\documentclass{article}
\usepackage{amsmath}
\usepackage{tabularx}
\usepackage{booktabs}
\usepackage{PRIMEarxiv}

\usepackage[utf8]{inputenc} % allow utf-8 input
\usepackage[T1]{fontenc}    % use 8-bit T1 fonts
\usepackage{hyperref}       % hyperlinks
\usepackage{url}            % simple URL typesetting
\usepackage{booktabs}       % professional-quality tables
\usepackage{amsfonts}       % blackboard math symbols
\usepackage{nicefrac}       % compact symbols for 1/2, etc.
\usepackage{microtype}      % microtypography
\usepackage{lipsum}
\usepackage{fancyhdr}       % header
\usepackage{graphicx}       % graphics
\graphicspath{{media/}}     % organize your images and other figures under media/ folder

\title{DA-TransUNet: Integrating Spatial and Channel Dual Attention with Transformer U-Net for Medical Image Segmentation
\thanks{\textit{{Corresponding Author}}: 
\textbf{Guanqun Sun, Le-Minh Nguyen, Junyi Xin}} 
}

\author{
  Guanqun Sun \\
  School of Information Engineering\\ 
  Hangzhou Medical College \\
  Hangzhou, Zhejiang, 311399, China\\
  School of Information Science\\
  Japan Advanced Institute of Science and Technology \\
  Nomi, Ishikawa, 923-1292, Japan\\
  \texttt{sun.guanqun@hmc.edu.cn} \\
   \And
  Yizhi Pan \\
  School of Information Engineering \\ 
  Hangzhou Medical College \\
  Hangzhou, Zhejiang, 311399, China\\
  \texttt{panyizhi@hmc.edu.cn} \\
   \And
  Weikun Kong \\
  Department of Electronic Engineering \\
  Tsinghua University \\
  Beijing, 100084, China \\
  \texttt{kong\_diison@mail.tsinghua.edu.cn} \\
   \And
  Zichang Xu \\
  Immunology Frontier Research Institute \\
  Osaka University \\
  Suita 565-0871, Japan\\
  \texttt{xuzichang@biken.osaka-u.ac.jp} \\
   \And
  Jianhua Ma \\
  Computer and Information Sciences \\
  Hosei University \\
  Tokyo184-8584, Japan\\
  \texttt{jianhua@hosei.ac.jp} \\
   \And
  Teeradaj Racharak, Le-Minh Nguyen \\
  School of Information Science\\
  Japan Advanced Institute of Science and Technology \\
  Nomi, Ishikawa, 923-1292, Japan\\
  \texttt{\{racharak, nguyenml\}@jaist.ac.jp} \\
   \And
  Junyi Xin \\
  School of Information Engineering\\ 
  Hangzhou Medical College \\
  Hangzhou, Zhejiang, 311399, China\\
  \texttt{ailab@hmc.edu.cn} \\
}

\begin{document}
\maketitle

\begin{abstract}
% \lipsum[1]
%part1: 目的: 背景, 问题冲突
Accurate medical image segmentation is critical for disease quantification and treatment evaluation. 
While traditional U-Net architectures and their transformer-integrated variants excel in automated segmentation tasks. 
However, they lack the ability to harness the image's intrinsic position and channel features.
%Specifically, they do not adequately capture features from both the spatial and channel dimensions inherent to image data. 
Existing models also struggle with parameter efficiency and computational complexity, often due to the extensive use of Transformers.
% Great progress has been made in automatic medical image segmentation due to powerful deep representation learning. 
% The influence of transformer has led to research into its variants, and large-scale replacement of traditional CNN modules. 
% However, such trend often overlooks the intrinsic feature extraction capabilities of the transformer and potential refinements to both the model and the transformer module through minor adjustments.
%part2本文的方法
To address these issues, this study proposes a novel deep medical image segmentation framework, called DA-TransUNet, aiming to integrate the Transformer and dual attention block(DA-Block) into the traditional U-shaped architecture. 
Unlike earlier transformer-based U-net models, DA-TransUNet utilizes Transformers and DA-Block to integrate not only global and local features, but also image-specific positional and channel features, improving the performance of medical image segmentation.
By incorporating a DA-Block at the embedding layer and within each skip connection layer, we substantially enhance feature extraction capabilities and improve the efficiency of the encoder-decoder structure.
% In contrast to previous transformer-based solutions, our DA-TransUNet leverages the multi-faceted feature extraction capabilities of transformers and DA-Block to effectively combine global, local, and image-specific spatial and channel characteristics, thereby enhancing medical image segmentation.
% Meanwhile, experimental results show that DA-Block added before the Transformer layer can facilitate feature extraction in the U-net structure. 
% Furthermore, incorporating DA-Block in skip connections can promote feature transfer to the decoder, thereby improving image segmentation performance.
%part3应用实验
DA-TransUNet demonstrates superior performance in medical image segmentation tasks, consistently outperforming state-of-the-art techniques across multiple datasets.
%Experimental results across various benchmark of medical image segmentation reveal that DA-TransUNet significantly outperforms the state-of-the-art methods.
%
%part4总结
In summary, DA-TransUNet offers a significant advancement in medical image segmentation, providing an effective and powerful alternative to existing techniques. Our architecture stands out for its ability to improve segmentation accuracy, thereby advancing the field of automated medical image diagnostics.
The codes and parameters of our model will be publicly available at https://github.com/SUN-1024/DA-TransUnet.
%由于强大的深度表示学习，自动医学图像分割取得了巨大的进展。受Transformer的影响，人们致力于研究Transformer的变体，和大规模的将传统CNN模块给替换掉。然而，这样子的行为忽略了transformer自身的特征提取和对Transformer提取能力的提升，同时忽略了小改动对于模型的提升和对于Transfomer模块的提升。在本文中，我们提出了一种新颖的深度医学图像分割框架，称为DA-TransUNet,旨在Transformer和DA-block引入到传统U型架构的编码器和解码器中。我们的DA-TransUNet利用了Transformer的自注意力机制和DA-Block多方面的特征提取，可以有效地结合全局，局部和多尺度的特征，从而增强医学图像分割能力。对医学图像分割几个经典的数据进行了实验，其结果都证明了DA-TransUNet的有效性。同时注重于开源，我们将我们模型的参数上传到了Hugging Face中。
\end{abstract}

% keywords can be removed

\section{Introduction}   
\label{sec:introduction}
Medical image segmentation is the process of delineating regions of interest within medical images for diagnosis and treatment planning. It serves as a cornerstone in medical image analysis.
Precise delineation of lesions plays a crucial role in quantifying diseases, aiding the assessment of disease prognosis, and evaluating treatment efficacy. Manual segmentation is both accurate and affordable for pathology diagnosis but vital in standardized clinical settings.
Conversely, automated segmentation ensures a reliable and consistent process, boosting efficiency, cutting down on labor and costs, and preserving accuracy. Consequently, there is a substantial demand for exceptionally accurate automated medical image segmentation technology within the realm of clinical diagnostics.
%病灶的精确勾画对于疾病的量化、疾病预后的评估以及治疗效果的评估起着至关重要的作用。 虽然以其准确性而闻名的专业手动分割对于一般病理诊断来说并不会过度占用资源或昂贵，但其应用在标准化临床环境中变得势在必行。 另一方面，自动分割提供了可靠且可重复的过程，可以提高效率、减少劳动力和费用并保持精度。 因此，临床诊断领域对极其准确的自动化医学图像分割技术有很大的需求。

%老问题段落
In the last decade, the traditional U-net structure have been widely employed in numerous segmentation tasks, yielding commendable outcomes. Notably, the U-Net model \cite{U-Net}, along with its various enhanced iterations, have achieved substantial success. ResUnet \cite{resunet} emerged during this period, influenced by the residual concept. Similarly, UNet++ \cite{unet++} emphasizes enhancements in skip connections, and DAResUnet \cite{daresunet} incorporates a dual attention block with a residual block (Res-Block) in U-net. Both architectures have benefited from the influence of the encoder-decoder idea, while skip connections provide initial features for the decoder, bridging the semantic gap between encoders and decoders. 
However, limitations in the sensing field and biases in convolutional operations can compromise segmentation accuracy. 
Additionally, the inability to establish long-range dependencies and global context further constrains performance improvements.

%开始说"新问题", UN下面一段缩短重写. Finished
The transformer \cite{attention}, originally developed for sequence-to-sequence modeling in Natural Language Processing (NLP), has also found utility in the field of Computer Vision (CV).
ViTs segment images into patches and input their embeddings into a transformer network for strong performance. \cite{ViT}.
The segmentation efficacy is amplified by ViTs' application in CV, especially in medical image segmentation.
Inspired by ViTs, TransUNet \cite{transunet} further combines the functionality of ViTs with the advantages of U-net in the field of medical image segmentation. 
Specifically, it employs a transformer's encoder to process the image and employs CNN and hopping connections for accurate up-sampling feature recovery, it neglects image-specific features like position and channel. 
Leveraging the capabilities of ViTs, TransUNet \cite{transunet}  fuses the strengths of ViTs and U-net architectures to advance the performance in medical image segmentation. TransUNet utilizes a transformer-based encoder for robust image feature extraction while incorporating conventional convolutional neural networks and skip connections to achieve precise feature map up-sampling. It omits considerations for image-specific attributes such as spatial positioning and channel information.
Swin-Unet \cite{swin-unet} combines the Swin-transform block with the U-net structure and achieves good results. Yet, adding extensive Transformer blocks inflates the parameter count without significantly improving results.
However, the aforementioned medical image segmentation studies show progress in leveraging U-net and Transformer features, they have some limitations:
%近十年来，卷积神经网络（CNN）被广泛应用于各种分割相关任务，并在医学影像分割领域取得了良好的成绩。近年来，优秀的全卷积网络（FCN[]）、U-Net[]及其各种改进模型都取得了巨大成功（ResUnet[]就是当时在残差思想影响下出现的模块之一，Unet++[]也是一种偏向于在跳接处进行改进的模型，Unet DAResUnet[]中还包含了DA块与Res块）。这两种架构都得益于编码器-解码器思想的影响，而跳转连接为解码器提供了初始特征，弥合了编码器和解码器之间的语义鸿沟。然而，由于底层模块中卷积运算的数量较多、卷积运算中出现的偏差以及无法建立长距离依赖关系和全局上下文等原因，传感领域受到了一些限制，这在一定程度上阻碍了分割精度的进一步提高。值得注意的是，在这种情况下，转换器应运而生[]，尽管转换器的首次应用是在自然语言处理（NLP）领域，设计用于模型中的序列到序列建模，但现在它已成为计算机视觉（CV）领域的一项技术。例如：视觉转换器（ViTs）[]通过将图像分割成多个小图像模块，然后将这些小图像模块的先验嵌入序列输入网络作为转换器的输入，取得了良好的效果。转换器的强大功能进一步提高了分割的准确性。受 ViT 的启发，TransUNet[] 进一步将 ViT 的功能与 U-Net 的优势结合起来，应用于医学图像分割领域。Swin-Unet[]将Swin变换块与Unet结构相结合，取得了良好的效果。在上述基于医学图像分割的工作中，有基于 Unet 特征的改进，也有基于变换器的改进。在某种程度上，上述工作令人鼓舞，但也存在一些局限性。
\begin{itemize} 
%TODO （总结性的提出limitations） DA名字的统一，一些DA Res的详细介绍在缩写。 over
% \item [1)]
%     In the traditional U-shaped medical image segmentation model, the extended convolution process results in a deficiency of global information, marking global feature acquisition challenging.
% %在传统U型医学图像分割模型中，由于长久的卷积过程，会存在全局信息的缺失，所以对于全局特征的获取一直都是一个问题。
    \item[1)]
    Although combining the Transformer with traditional U-Net architectures has shown promise in medical image segmentation, the Transformer lacks built-in mechanisms for considering image-specific features of position and channel. This gap in functionality calls for additional investigation.
    % While the fusion of Transformer with the traditional U-Net has yielded promising results in medical image segmentation, Transformer does not inherently consider the characteristics of the image (channel and position) for feature extraction. This is an area that warrants further research.
    % With the introduction of the attention mechanism, numerous attention modules have been added to U-net. However, the challenge lies in optimally positioning and utilizing these attention modules to aid in feature extraction.
%随着注意力机制的引入，U-net中添加了大量的注意力模块。 然而，挑战在于最佳定位和利用这些注意模块来帮助特征提取。
%From SUN: 可否也换个角度, 目前对如何在U结构上充分挖掘利用Attention机制, 没有根据图像的特点来结合注意力机制, channel的问题 还不够深入, 还有......的问题.
    \item[2)] 
    In the U-Net model, skip connections serve as a vital element, bridging the semantic divide between the encoder and the decoder. Despite their potential to improve segmentation performance, skip connections have seen limited optimization efforts to date.
    % Skip connections are a crucial component in the U-Net model, bridging the semantic gap between the encoder and the decoder. However, so far, there has been limited exploration into optimizing skip connections, despite their potential to enhance the model's segmentation capabilities.
    % Transformer is powerful in extracting features while being able to retain global information. However, further stimulate the potential of Transformer and how to effectively combine the features of both CNN and Transformer is a question worth thinking about.

    % The Transformer exhibits robust feature extraction capabilities while retaining global context. However, smoothly transitioning from CNN to Transformer and fully unlocking the potential of the Transformer, effectively amalgamating the strengths of both CNN and Transformer, presents a thoughtful challenge.
%Transformer提取特征能力强大，同时能够保留全局信息。但是如何将Transformer的潜能进一步激发出来，如何将CNN和Transformer两者的特点有效地结合是一个值得思考的问题。
%From SUN: 如果说这个是当前的limitaion的问题, 可以换个角度来说: transformer在U-net架构中的挖掘还不够, 都是堆叠多个trans,导致参数量过大, 效果提升不大, 没有充分发挥.   
    \item[3)] 
    Many studies merely stack multiple Transformers to enhance models, resulting in inflated parameters and computational complexity with marginal gains in performance. The intricate design of integrating Transformers and U-Net architectures warrants further investigation.
    %Transformer has powerful feature extraction capabilities and global information extraction capabilities. The Transformer is not explored deeply enough in the U-net architecture. Multiple Transformers are stacked, resulting in too large parameters but little improvement in effect. This idea is not feasible.
%跳跃连接在Unet模型中是一个十分重要的组成部分，弥补经过编码器到解码器的语义鸿沟，但是如何帮助解码器获得更精准的特征以帮助解码器返还出更精准地特征原图，进一步提高模型鲁棒性仍然是存在问题。
\end{itemize}

To address the aforementioned challenges, we propose DA-TransUNet, which incorporates DA-Blocks specifically designed to extract image-specific positional and channel features, thereby enhancing both parameter efficiency and performance.
We believe that the extensive use of Transformers is not as impactful as utilizing a suite of precisely calibrated DA-Blocks specifically optimized for image-specific features.
The DA-Block within the transformer layer possesses robust, specialized capability for extracting image-specific positional and channel features.
This block integrates the Position Attention Block (PAM) and Channel Attention Block (CAM) from the Dual Attention Network for Scene Segmentation\cite{dual01}. Positioned in the embedding layer of DA-TransUNet, the Dual Attention Block offers robust feature extraction capabilities.
We also integrate DA-Block into the three-layer skip connection to optimize features passed by the encoder. This narrows the semantic gap and aids in creating a unified feature representation.
This fusion method maximizes the use of positional and channel features in the attention mechanism, optimizing the model.
Furthermore,  skip connections in the U-shape structure are enhanced with DA-Blocks to filter irrelevant information, improving image reconstruction quality.
% Unlike traditional Unet-based encoders, we incorporate a Transformer for enhanced distant contextual feature extraction, boosting encoding capabilities. To improve the decoder, we refine features via skip connections.
Owing these enhancements, both the decoding and medical image segmentation capabilities are significantly bolstered. 
% We evaluated the effectiveness of DA-TransUNet using the CVC-ClinicDB dataset\cite{cvc-clinicDB}, the Kvasir SEG dataset\cite{kvasir-Seg}, the Kvasir-Instrument dataset\cite{Kvasir-Instrument}, the Synapse dataset\cite{Synapse} and Chest X-ray mask and label dataset\cite{jaeger2013automatic,candemir2013lung}.
%在上述的问题和局限性中，我们提出了我们的模型DA-TransUnet。我们觉得大规模的使用Transformer起到的作用不如仅仅使用一个来的作用要打大，所以我们在TransUNet的底子上进行了改进。为了进一步为transformer提供更加精准的特征，我们借用了 Dual Attention Network for Scene Segmentation（场景分割双注意力网络）中的位置注意力块（PAM）和通道注意力块（CAM）组成了一个双重注意力块（DA-Block），将其加在了嵌入层中，为transformerlayer 提供了更精准的特征。考虑到编码器解码器和跳跃连接的结构，我们在三层跳跃连接中对编码器传递的特征使用DA-Block进行特征优化，帮助解码器编码器结构进一步减少语义鸿沟，以生成统一的特征表征。这种融合方法可以充分的挖掘不同方面的、全局和局部的特征，并使他们相互补充。与传统的U型结构一致，使用编码器获取上下文特征，并采用跳跃连接和解码器进行统一特征的融合。与基于CNN的大多数编码器不同的是，我们在编码中加入了Transformer模块，来进一步提取远距离的上下文特征，从而大大提升编码能力，同时为了提高解码器的能力，我们将跳跃连接所传递的特征进一步进行提炼。受益于这些改进，解码能力提升，医学图像分割能力也在加强。我们使用了 CVC-ClinicDB 数据集、Kvasir SEG 数据集、Kvasir-Instrument 数据集和 Synapse 数据集和Chest X-ray mask and label dataset数据集，对 DA-TransUNet 的有效性进行了评估。

We mainly evaluate the effectiveness of proposed DA-TransUNet on several medical image datasets of Synapse \cite{Synapse}, CVC-ClinicDB \cite{cvc-clinicDB}, ISIC2018 \cite{ISIC01,ISIC02}, kvasir-seg \cite{kvasir-Seg}, Kvasir-Instrument dataset \cite{Kvasir-Instrument} and Chest X-ray mask and label dataset \cite{jaeger2013automatic,candemir2013lung}. DA-TransUNet demonstrates notable efficacy, as evidenced by quantifiable metrics. Our main contributions are summarized below:
%在实验中经多个数据集验证，DA-TransUNet都有不错的成功率。我们的主要贡献总结如下：
\begin{itemize}
    \item [1)]
    We propose DA-TransUNet, a novel architecture that integrates dual attention mechanisms to process positional and channel information into a Transformer U-net framework. This design improves the flexibility and functionality of the encoder-decoder structure, thereby improving performance in medical image segmentation tasks.
    % We propose DA-TransUNet, a novel architecture that effectively combines dual attention mechanisms of position and channel with Transformer components for medical image segmentation. which enhances the functionality and flexibility of the encoder-decoder structure, and further improves the modeling of medical image segmentation. 有待润色到一起
    %我尝试一下更加简介的介绍contribution, 把之前的版本注释. 
    % In this paper, we propose a framework based on the dual-attention mechanism and the transformer, called DA-TransUNet, which enhances the functionality and flexibility of the encoder-decoder structure, and further improves the modeling of medical image segmentation.
    %本文提出了一种基于双注意力机制和变换器的框架，称为 DA-TransUNet，它有效地结合了变换器的优势，增强了编码器-解码器结构的功能性和灵活性，进一步提高了医学图像分割的建模效果。
    \item [2)]
    A well-designed dual-attention encoding mechanism is proposed to be positioned ahead of the Transformer layer in encoder. This can enhance its feature extraction capabilities and enrich the functionality of the encoder in the U-net structure.(Section \ref{Method})
    % We propose a new perspective to enhance the feature extraction capability of Transformer, i.e., placing DA-Block in front of the Transformer layer to provide Transformer with more accurate feature maps, in order to improve the feature extraction capability of Transformer and to enhance the functionality of the face encoder.
    %第二点要不要更简介一点, 主要就是: DA块来增强transformer的提取能力
    %我们提出了增强Transformer特征提取能力的新视角，即将DA-Block放在Transformer层前面，提供给Transformer更准确的特征图，以提高Transformer的特征提取能力，增强面编码器的功能。
    \item[3)] 
    We enhance the effectiveness of skip connections by incorporating Dual Attention Block into each layer, a modification substantiated by ablation studies, which results in more accurate feature delivery to the decoder and improved image segmentation performance.(Section \ref{ablation})
    % We propose a new perspective to enhance the effectiveness of features delivered by jump connections, i.e., adding DA-Block to each layer in the jump connection to provide more accurate features for the decoder to help the decoder recover the original feature maps and improve the model's image segmentation capability.
    %我们提出了增强跳跃连接传递特征有效性的新视角，即将DA-Block加入到跳跃连接中的每一层，为解码器提供更为准确的特征，以帮助解码器恢复原始特征图，提高模型的图像分割能力。
    \item[4)]
    Our proposed DA-TransUNet method achieves state-of-the-art performance on multiple medical imaging datasets, which proves the effectiveness of our method and its contribution to advcancing medical image segmentation.
    %在Synapse[]，CVC-ClinicDB[],ISIC2018[],kvasir-seg[]等六个数据集中，DA-TransUNet方法都得到了不错的成功率，证明了我们方法有效性和优越性。
    % \item[5)]
    % We also uploaded the model trained parameters, to Hugging Face.
    %我们还将模型训练好的参数，上传到了Hugging Face上。
\end{itemize}
The rest of this article is organized as follows. Section II reviews the related works of automatic medical image segmentation, and the description of our proposed DA-TransUNet is given in Section III. Next, the comprehensive experiments and visualization analyses are conducted in Section IV. Finally, Section V makes a conclusion of the whole work.
%本文接下来的内容安排如下。第二节回顾了医学影像自动分割的相关工作，第三节介绍了我们提出的 DA-TransUNet 技术。接下来，第四部分进行了全面的实验和可视化分析。最后，第五部分对整个工作进行总结。

\section{Related Work}
\subsection{U-net Model}
Recently, attention mechanisms have gained popularity in U-net architectures\cite{U-Net}. For example, Attention U-net incorporates attention mechanisms to enhance pancreas localization and segmentation performance\cite{attention-unet}; DAResUnet integrates both double attention and residual mechanisms into U-net \cite{daresunet}; Attention Res-UNet explores the substitution of hard-attention with soft-attention \cite{attention-res-unet}; Sa-unet incorporates a spatial attention mechanism in U-net\cite{sa-unet}. Following this, TransUNet innovatively combines Transformer and U-net structure \cite{transunet}. Building on TransUNet, TransU-Net++ incorporates attention mechanisms into both skip connections and feature extraction\cite{transunet++}. Swin-Unet\cite{swin-unet} improves by replacing every convolution block in U-net with Swin-Transformer\cite{swin-Transformer}. DS-TransUNet proposes to incorparte the tif module to the skip connection to improve the model\cite{ds-transunet}. AA-transunet leverages Block Attention Model (CBAM) and Deep Separable Convolution (DSC) to further optimize TransUNet \cite{aa-transunet}. TransFuse uses dual attention Bifusion blocks and AG to fuse features of two different parts of CNN and Transformer\cite{transfuse}. Numerous attention mechanisms have been added to U-net and TransUNet models, yet further exploration is warranted. Diverging from prior approaches, our experiment introduces a dual attention mechanism and Transformer module into the traditional U-shaped encoder-decoder and skip connections, yielding promising results.

\subsection{Application of skip connections in medical image segmentation modeling}
Skip connections in U-net aim to bridge the semantic gap between the encoder and decoder, effectively recovering fine-grained object details \cite{drozdzal}\cite{he}\cite{huang}. There are three primary modifications to skip connections: firstly, increasing their complexity \cite{unet-review}. U-Net++ redesigned the skip connection to include a Dense-like structure in the skip connection\cite{unet++}, and U-Net3++\cite{unet3++} changed the skip connection to a full-scale skip connection. Secondly, RA-UNet introduces a 3D hybrid residual attention-aware method for precise feature extraction in skipped connections \cite{ra-unet}. The third is a combination of encoder and decoder feature maps: An alternative extension to the classical skip connection was introduced in BCDU-Net with a bidirectional convolutional long-term-short-term memory (LSTM) module was added to the skip connection\cite{bcdu-net}. Aligning with the second approach, we integrate Dual Attention Blocks into each skip connection layer, enhancing decoder feature extraction and thereby improving image segmentation accuracy.
%跳转连接机制在 UNet 中被使用，其目的是弥合编码器和解码器之间的语义鸿沟，并已被证明能够有效地恢复目标对象的细粒度细节。对跳转连接进行修改的方式主要有三种，第一种是增加跳转连接的复杂度：U-Net++重新设计了跳转连接，在跳转连接中加入了类似Dense的结构（Dense-like structure）；U-Net3++则将跳转连接改为全面跳转连接；第二种是对跳转连接中的特征图进行处理： RA-UNet提出了一种三维混合残留注意力感知分割方法，用于精确提取图像特征。第三种是编码器和解码器特征图的组合： 在 BCDU-Net 中引入了对经典跳转连接的另一种扩展，在跳转连接中加入了双向卷积长短期记忆（LSTM）模块。与第二种想法一致（在跳过连接中处理特征图）我们在跳跃连接的每一层中都加入了DA-Block，来帮助解码器从编码器获取更加精准的特征，以提高图像分割的准确率。

\subsection{The use of attentional mechanisms in medical images}
Attention mechanisms are essential for directing model focus towards relevant features, thereby enhancing performance. In recent years,  dual attention mechanisms have seen diverse applications across multiple fields. In scene segmentation, the Dual Attention Network (DANet) employs position and channel attention mechanisms to improve performance \cite{dual01}.
A modularized DANs framework is presented that adeptly merges visual and textual attention mechanisms \cite{dual02}. This cohesive approach enables selective focus on pivotal features in both types of data, thereby improving task-specific performance.
Additionally, the introduction of the Dual Attention Module (DuATM) has been groundbreaking in the field of audio-visual event localization. This model excels at learning context-aware feature sequences and performing attention sequence comparisons in tandem, effectively incorporating auditory-oriented visual attention mechanisms \cite{dual04}.
Moreover, dual attention mechanisms have been applied to medical segmentation, yielding promising results\cite{dual05}. The Multilevel Dual Attention U-net for Polyp Segment combines dual attention and U-net in medical image segmentation\cite{mlda-unet}. While significant progress has been made in medical image segmentation, there is still ample room for further research to explore the potential of position and channel attention mechanism in the field of medical image segmentation.
%注意力机制是帮助模型聚焦目标特征并提高性能的关键组成部分。 近年来，双重注意机制在各个领域得到了广泛的应用。
% 在场景分割领域，引入了用于场景分割的双注意力网络（DANet），它通过引入空间和通道注意力机制来增强场景分割的性能\cite{dual01}。
% 在论文“Dual Attention Networks for Multimodal Reasoning and Matching”中，提出了模块化的 DAN 框架。 该框架有效地结合了视觉和文本注意力，以关注给定任务的视觉和文本数据的关键部分\cite{dual02}。
% 此外，论文《Dual Attention Matching for Audio-Visual Event Localization》介绍了一种面向听觉的视觉注意模型。 此外，提出了双重注意力模块（DuATM）来学习上下文感知特征序列并同时执行注意力序列比较\cite{dual04}。
% 此外，双重注意机制已应用于医学分割，产生了有希望的结果\cite{dual05}。 Multilevel Dual Attention Unet for Polyp Segment 将双重注意力和 Unet 结合在医学图像分割中\cite{mlda-unet}。 虽然医学图像分割已经取得了重大进展，但仍有足够的空间进行进一步研究，以探索双重注意力块在医学图像分割领域的潜力。
\section{Method}
\label{Method}
%Method中的每个图的描述语言, 需要更加详细一点, 具体可以参考我们的重点关注的几篇例文 over
\begin{figure*}[t!]
\centering
\includegraphics[width=\textwidth]{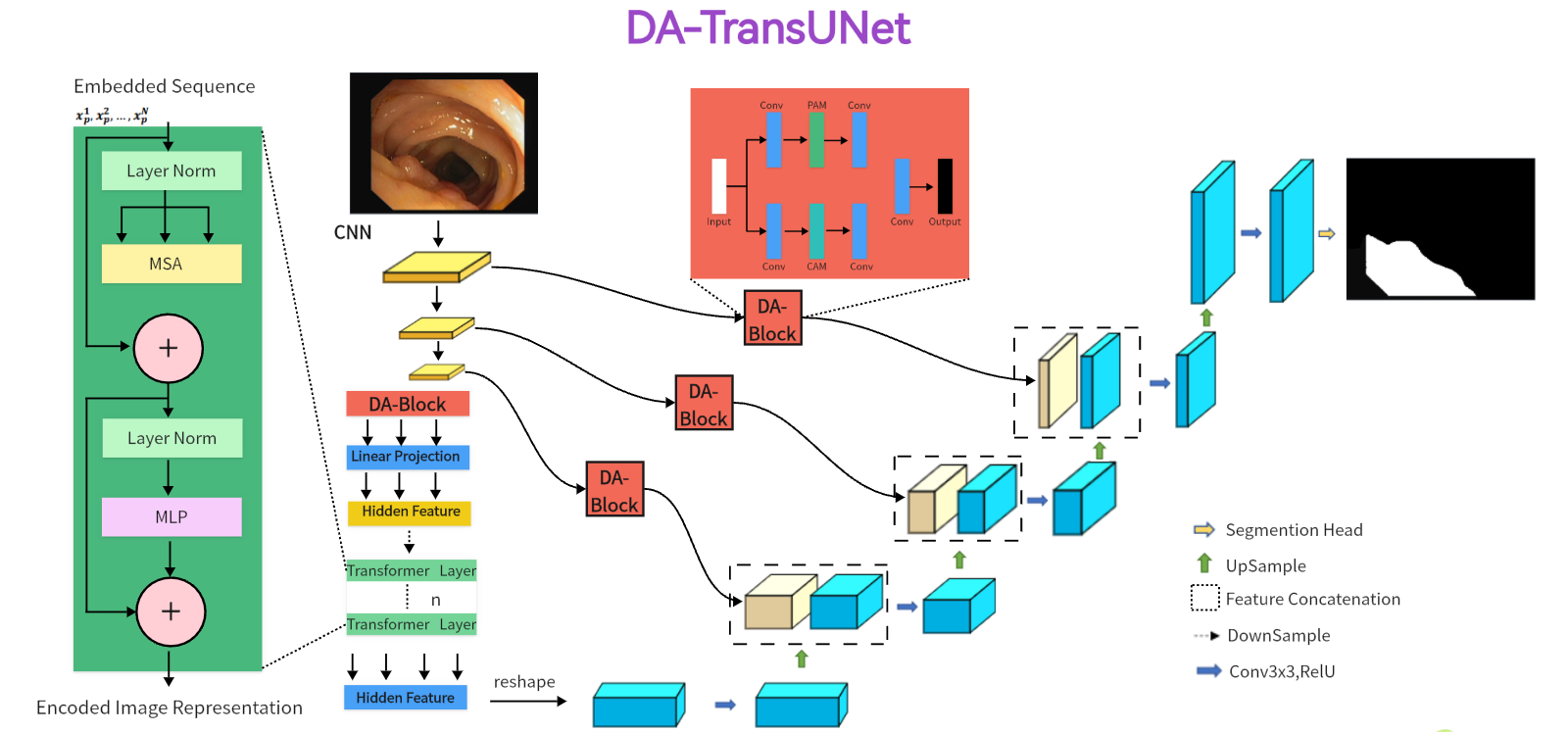}
\caption{\textbf{Illustration of the proposed dual attention transformer U-Net(DA-TransUNet). For the input medical images, we feed them into an encoder with transformer and Dual Attention Block (DA-Block). Then, the features of each of the three different scales are purified by DA-Block. Finally, the purified skip connections are fused with the decoder, which subsequently undergoes CNN-based up-sampling to restore the channel to the same resolution as the input image. In this way, the final image prediction result is obtained.}}
%拟议的双注意变换器 U-Net（DA-TransUNet）示意图。对于输入的医学图像，我们将其输入带有transformer和DA-Block的编码器。然后，将三层不同的尺度的特征分别经过DA-Block提纯。最后，将提纯后的跳跃连接与解码器融合，随后经过基于CNN的上采样处理，恢复道与输入图像相同的分辨率。这样，得到最后的图像预测结果。
\label{Fig_DATrans-Unet}
\end{figure*}
In the subsequent section, we propose the DA-TransUNet architecture, illustrated in Figure.\ref{Fig_DATrans-Unet}. We start with a comprehensive overview of the architecture. Next, we detailed the architecture's key components in the following order: the dual attention blocks(DA-Block), the encoder, the skip connections, and the decoder.
\subsection{Overview of DA-TransUNet}
In Figure \ref{Fig_DATrans-Unet}, the architecture of DA-TransUNet is presented. The model comprises three core components: the encoder, the decoder, and the skip connections. In particular, the encoder fuses a conventional convolutional neural network (CNN) with a Transformer layer and is further enriched by the DA-Block, which are exclusively introduced in this model architecture. In contrast, the decoder primarily employs conventional convolutional mechanisms. 
For the optimization of skip connections, DA-Blocks serve as pivotal components within the DA-TransUNet architecture.
DA-Blocks filter irrelevant information in skip connections, enhancing image reconstruction accuracy.
In summary, in contrast to traditional convolutional approaches and the extensive use of Transformers, DA-TransUNet uniquely leverages DA-Blocks for the extraction and utilization of image-specific features of position and channel. This strategic incorporation significantly elevates the overall performance of the model.

To elucidate the rationale behind our proposed DA-TransUNet model's design, it's imperative to consider the limitations and strengths of both U-Net architectures and Transformers in the context of feature extraction. While Transformers excel in global feature extraction through their self-attention mechanisms, they are inherently limited to unidirectional focus on positional attributes, thus neglecting multi-faceted feature perspectives. On the other hand, traditional U-Net architectures are proficient in local feature extraction but lack the capability for comprehensive global contextualization. To address these constraints, we integrate DA-Blocks both preceding the Transformer layers and within the encoder-decoder skip connections. 
This achieves two goals: firstly, it refines the feature map input to the Transformer, enabling more nuanced and precise global feature extraction; secondly, the DA-Block in the skip connections optimize the transmitted features from the encoder, facilitating the decoder in reconstructing a more accurate feature map. Thus, our proposed architecture amalgamates the strengths and mitigates the weaknesses of both foundational technologies, resulting in a robust system capable of image-specific feature extraction.

\subsection{Dual Attention Block(DA-Block)}
\begin{figure*}[t!]
\centering
\includegraphics[width=0.8\textwidth]{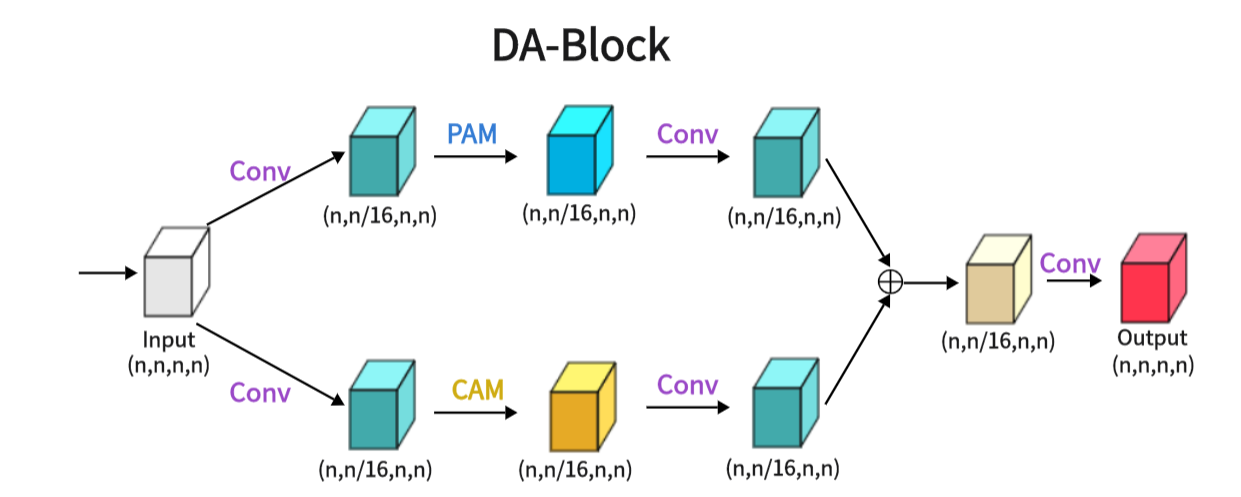}
\caption{\textbf{The proposed Dual Attention Block (DA-Block) is shown in the Figure. The same input feature map is input into two feature extraction layers, one is the position feature extraction block and the other is the channel feature extraction block, and finally, the two different features are fused to obtain the final DA-Block output.}}
\label{Fig_DA}
\end{figure*}

%整体上来看现在是介绍DA比较细致了, 介绍了是什么, 但其实更加详细的应该是介绍为什么PAM,CAM的DA在我们需要用到, 也就是DA的优点好处(要和我们的医学分割任务结合说优点好处) over
%-------> 介绍PAM, CAM的篇幅可以减少, 把这部分篇幅让给他们的优点. 或者说详细描述PAM和CAM的时候, 融入讲他们的优点 , 结合在一起.  over
As shown in the attached Figure \ref{Fig_DA}, the Dual Attention Block (DA-Block) serves as a feature extraction module that integrates image-specific features of position and channel. This enables feature extraction tailored to the unique attributes of the image. 
Particularly in the context U-Net shaped architectures, the specialized feature extraction capabilities of the DA-Block are crucial. While Transformers are adept at using attention mechanisms to extract global features, they are not specifically tailored for image-specific attributes. In contrast, the DA-Block excels in both position-based and channel-based feature extraction, enabling a more detailed and accurate set of features to be obtained.
% While the Transformer possesses excellent feature extraction capabilities, focusing on different positions, it may overlook the requirement for precision in medical images. In contrast, the DA-Block excels not only in position-based feature extraction but also in channel-based feature extraction. 
% Compared to single-feature extraction, multi-angle feature extraction can acquire more accurate and detailed features, thereby improving the model's segmentation capabilities. 
%大小前提的结论
Therefore, we incorporate it into the encoder and skip connections to enhance the model's segmentation performance. The DA-Block consists of two primary components: one featuring a Position Attention Module (PAM), and the other incorporating a Channel Attention Module (CAM), both borrowed from the Dual Attention Network for scene segmentation\cite{dual01}.
% TODO: 上面的DA还需要一句话的定义, 是PAM和CAM的组合. (需要一句定义来说是什么) over
% TODO: 说说为什么需要DA? 只是trans的优点是没有遗忘, Attention可以关注到不同位置的特征, 缺点是忽视了医学图像的特点需要更加精确.  前提:图像如果两个角度分离提出特征更加精准关注更加细致, 因此本文中的DA则可以不仅从Position还从channel的角度来提取可以更多特征. (下面再说结构) over

%这里还需要带着说一下, DA在DA-TransUnet中encoder和skip中都用到了. 再说DA由PAM,CAM构成. 再简略的提一句为什么需要用DA->可以提取更深入的特征. (等等后面介绍了DA之后(DA是什么), 再详细的说DA的好处是什么(为什么用DA). 
% DA=PAM+CAM是谁提出来的, 需要加一下引用, 说明一下. over

% Let me start with the PAM and CAM introduced by Dual Attention Network for Scene Segmentation.(可以去掉, 上面已经介绍2部分了, 学术论文中一般不用let me, 直接从PAM开始介绍就行) over
% PAM ........加黑:  over

\begin{figure}[t!]
\centering
\includegraphics[width=0.7\columnwidth]{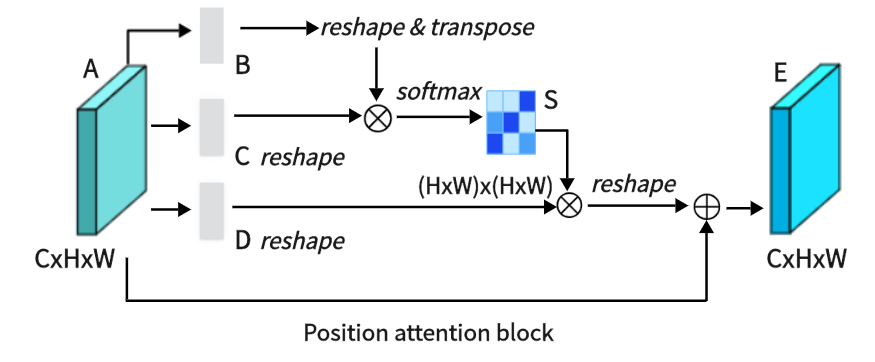}
\caption{\textbf{Architecture of Position Attention Mechanism(PAM).}}
\label{Fig_PAM}
\end{figure}

\textbf{PAM (Position Attention Module)}: As shown in Figure \ref{Fig_PAM}, PAM captures spatial dependencies between any two positions of feature maps, updating specific features through a weighted sum of all position features. 
The weights are determined by the feature similarity between two positions. Therefore, PAM is effective at extracting meaningful spatial features.

PAM initially takes a local feature, denoted as A $\in$ $R^{C\times H\times W}$ (C represents Channel, H represents height, and W represents Width). We then feed A into a convolutional layer, resulting in three new feature maps, namely B, C, and D, each of size $R^{C\times H\times W}$. Next, we reshape B and C to $R^{C\times N}$, where N = H×W denotes the number of pixels. We perform a matrix multiplication between the transpose of C and B and subsequently use a softmax layer to compute the spatial attention map S $\in$ $R^{N\times N}$:
\begin{equation}
      S_{j i}=\frac{\exp \left(B_{i} \cdot C_{j}\right)}{\sum_{i=1}^{N} \exp \left(B_{i} \cdot C_{j}\right)}
\end{equation}
Here, $S_{ji}$ measures the impact of the i-th position on the j-th position. We then reshape matrix D to $R^{C\times N}$. A matrix multiplication is performed between D and the transpose of S, followed by reshaping the result to $R^{C\times H\times W}$. Finally, we multiply it by a parameter $\alpha$ and perform an element-wise sum operation with the features A to obtain the final output E $\in$ $R^{C\times H\times W} $:
\begin{equation}
      E_{j}=\alpha \sum_{i=1}^{N}(s j i D i)+A j
\end{equation}
The weight $\alpha$ is initialized as 0 and is learned progressively. PAM has a strong capability to extract spatial features. As E is generated as a weighted sum of all position features and original features, it possesses global contextual features and aggregates context based on the spatial attention map. This ensures effective extraction of position features while maintaining global contextual information.

\begin{figure}[t!]
\centering
\includegraphics[width=0.7\columnwidth]{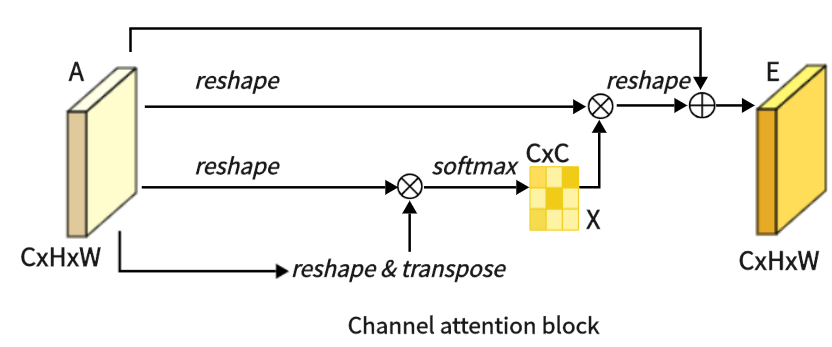}
\caption{\textbf{Architecture of Channel Attention Mechanism(CAM).}}
\label{Fig_CAM}
\end{figure}

\textbf{CAM (Channel Attention Module)}: As shown in Figure \ref{Fig_CAM}, this is CAM, which excels in extracting channel features. Unlike PAM, we directly reshape the original feature A $\in$ $R^{C\times H\times W}$ to $R^{C\times N}$, and then perform a matrix multiplication between A and its transpose. Subsequently, we apply a softmax layer to obtain the channel attention map X $\in$ $R^{C\times C}$:
\begin{equation}
      x_{j i}=\frac{\exp \left(A_{i} \cdot A_{j}\right)}{\sum_{i=1}^{C} \exp \left(A_{i} \cdot A_{j}\right)}
\end{equation}
Here, $x_{ji}$ measures the impact of the i-th channel on the j-th channel. Next, we perform a matrix multiplication between the transpose of X and A, reshaping the result to $R^{C\times H\times W}$. We then multiply the result by a scale parameter $\beta$ and perform an element-wise sum operation with A to obtain the final output E $\in$ $R^{C\times H\times W}$:
\begin{equation}
      E_{j}=\beta \sum_{i=1}^{N}(x j i A i)+A j
\end{equation}
Like $\alpha$, $\beta$ is learned through training. Similar to PAM, during the extraction of channel features in CAM, the final feature for each channel is generated as a weighted sum of all channels and original features, thus endowing CAM with powerful channel feature extraction capabilities.

\textbf{DA (Dual Attention Module)}: As shown in the Figure\ref{Fig_DA}, we present the architecture of the Dual Attention Block (DA-Block). This architecture merges the robust position feature extraction capabilities of the Positional Attention Module (PAM) with the channel feature extraction strengths of the Channel Attention Module (CAM). Furthermore, when coupled with the nuances of traditional convolutional methodologies, the DA-Block emerges with superior feature extraction capabilities. DA-Block consists of two components, the first one is dominated by PAM and the second one is dominated by CAM.
The first component takes the input features and performs one convolution to scale the number of channels by one-sixteenth to get $\alpha ^{1} $. This simplifies feature extraction by PAM; following a PAM feature extraction and another convolution,  $\hat{\alpha ^{1} } $ is obtained.
\begin{equation}
\alpha ^{1} = Conv\left ( input \right )
\end{equation}
\begin{equation}
\hat{\alpha ^{1} } = Conv\left ( PAM\left ( \alpha ^{1} \right ) \right )
\end{equation}
The other component is the same, with the only difference being that the PAM block is replaced with a CAM with the following formula:
\begin{equation}
\alpha ^{2} = Conv\left ( input \right )
\end{equation}
\begin{equation}
\hat{\alpha ^{2} } = Conv\left ( CAM\left ( \alpha ^{2} \right ) \right )
\end{equation}
After extracting $\hat{\alpha ^{1} } $ and $\hat{\alpha ^{2} } $ from the two layers of attention, the output is obtained by aggregating and summing the two layers of attention and recovering the number of channels in one convolution.
\begin{equation}
output = Conv\left ( \hat{\alpha ^{1} } +\hat{\alpha ^{2} } \right )
\end{equation}
This sophisticated DA-Block architecture seamlessly integrates the strengths of the PAM and CAM to improve feature extraction, making it a critical component in enhancing the model's overall performance.
% 如图所示，这是双重注意力块的架构，接下来我们将对它进行详细讲解。DA-Block由两层组成，第一层以PAM为主，第二层以CAM为主。
% 第一层先将输入特征进行一次卷积将通道数缩放十六分之一得到$\alpha ^{1} $，可以帮助PAM块更加轻松的提取特征，随后经过一次PAM块的特征提取和一次普通的卷积得到$\hat{\alpha ^{1} } $
% \begin{equation}
% \alpha ^{1} = Conv\left ( input \right )
% \end{equation}

% \begin{equation}
% \hat{\alpha ^{1} } = Conv\left ( PAM\left ( \alpha ^{1} \right ) \right )
% \end{equation}
% 另一层也是如此，唯一的区别是将PAM块替换成了CAM，公式如下：
% \begin{equation}
% \alpha ^{2} = Conv\left ( input \right )
% \end{equation}

% \begin{equation}
% \hat{\alpha ^{2} } = Conv\left ( CAM\left ( \alpha ^{2} \right ) \right )
% \end{equation}
% 将经过两层注意力得提取后的$\hat{\alpha ^{1} } $ 和 $\hat{\alpha ^{2} } $进行聚合求和后，在经过一次卷积恢复通道数得到输出。
% \begin{equation}
% output = Conv\left ( \hat{\alpha ^{1} } +\hat{\alpha ^{2} } \right )
% \end{equation}
% 这个复杂的DA模块架构无缝整合了位置注意模块和通道注意模块的优势，以改善特征提取，使其成为提高模型整体性能的关键组成部分。
% $\alpha ^{1} = Conv\left ( input \right )$
% $\alpha ^{2} = Conv\left ( input \right )$
% $\hat{\alpha ^{2} } $
% $\hat{\alpha ^{1} } $
% $\alpha ^{2} $
% $\alpha ^{1} $
\subsection{Encoder with Transformer and Dual Attention}
% TODO 123标出来 然后开始的  段首句说明。
As illustrated in Figure \ref{Fig_DATrans-Unet}, the encoder architecture consists of four key components: convolution blocks, DA-Block, embedding layers, and transformer layers. Of particular significance is the inclusion of the DA block before the Transformer layer. This design is aimed at performing specialized image processing on the post-convolution features, enhancing the Transformer's feature extraction for image content. While the Transformer architecture plays a crucial role in preserving global context, the DA block strengthens the Transformer's capability to capture image-specific features, enhancing its ability to capture global contextual information in the image. This approach effectively combines global features with image-specific spatial and channel characteristics.

%Particularly crucial is the inclusion of a DA-Block before the Transformer layer, which purifies the features obtained through three convolutional blocks from two different perspectives before feeding them into the Transformer. This enhances the feature extraction capability of the Transformer.
%TODO: 讲讲为什么encode中加入DA(我们提出来的), over
%下面再详细描述over
The first component comprises the three convolutional blocks of the architecture of the U-Net and its diverse iterations, seamlessly integrating convolutional operations with downsampling processes. Each convolutional layer halves the size of the input feature map and doubles its dimension, a configuration empirically found to maximize feature expressiveness while maintaining computational efficiency. 
The second component uses DA-Block extract features at both positional and channel levels, enhancing the depth of feature representation while preserving the intrinsic characteristics of the input map. 
The third component is the embedding layer serves as a critical intermediary, enabling the requisite dimensional adaptation, a prelude to the subsequent Transformer strata. 
The fourth component integrates Transformer layers for enhanced global feature extraction, beyond the reach of traditional CNNs.
% 以上的部分，在一起后encoder
Putting the above parts together, it works as follows: the input image traverses three consecutive convolutional blocks, systematically expanding the receptive field to encompass vital features. Subsequently, the DA-Block refines features through the application of both position-based and channel-based attention mechanisms. Following this, the remodeled features undergo a dimensionality transformation courtesy of the embedding stratum before they are channeled into the Transformer framework for the extraction of all-encompassing global features. This orchestrated progression safeguards the comprehensive retention of information across the continuum of successive convolutional layers. Ultimately, the feature map generated by the Transformer is restructured and guided through intermediate strata en route to the decoder.

% The Transformer architecture plays a cardinal role in the preservation of global context, while the DA block empowers the Transformer's feature acquisition, amplifying its adeptness in capturing potent global contextual information. This synergistic approach harmoniously amalgamates both global and localized features, effectively harmonizing their integration.%这一段可以放到encoder开头
% amag换词
By combining convolutional neural networks, transformer architectures, and dual-attention mechanisms, the encoder configuration culminates in a robust capability for feature extraction, resulting in a symbiotic powerhouse of capabilities.
%如图 \ref{Fig_DATrans-Unet} 所示，编码器架构由三个关键组件组成：三个卷积块、双注意力模块、嵌入层和专用于 Transformer 的层。 特别重要的是在 Transformer 层之前包含一个 DA-Block，它从两个不同的角度纯化通过三个卷积块获得的特征，然后将其输入 Transformer。 这增强了 Transformer 的特征提取能力。 三个组卷积块的组成符合 U-Net 的架构及其多样化的迭代，将卷积运算与下采样过程无缝集成。 每个卷积层都会将输入特征图的大小减半，并使其维度增加一倍。 双注意力模块执行特征增强和内在特征图保留的双重作用。 嵌入层充当关键的中介，实现必要的尺寸适应，这是后续 Transformer 层的前奏。 Transformer 层的同化为模型提供了熟练的全局特征提取能力，并确保整体信息的保存。
%在编码器划分的范围内，输入图像遍历三个连续的卷积块，系统地扩展感受野以包含重要特征。 随后，DA-Block 通过应用基于位置和基于通道的注意力机制来细化特征。 接下来，重构的特征在被导入 Transformer 框架以提取无所不包的全局特征之前，会经过嵌入层的维度转换。 这种精心安排的进程保证了连续卷积层连续体中信息的全面保留。 最终，由 Transformer 生成的特征图被重组并引导通过中间层到达解码器。
%Transformer 架构在保存全局上下文方面发挥着重要作用，而 DA 块则增强了 Transformer 的特征获取能力，增强了其捕获有效全局上下文信息的能力。 这种协同方法和谐地融合了全球和本地特征，有效地协调了它们的整合。
%通过合并传统的卷积神经网络 (CNN)、Transformer 架构和双注意力机制，编码器配置最终实现了强大的特征提取能力，从而形成了强大的共生能力。
\subsection{Skip-connections with Dual Attention}
%TODO: 先说我们的skip是什么: 本文创新提出了加入了DA. 再说为什么加入DA(有效过滤). 再结合还原图像说加入DA的原因. 可以说和传统是一样的.  over
Similar to other U-structured models, we have also incorporated skip connections between the encoder and decoder to bridge the semantic gap that exists between them. 
To further minimize this semantic gap, we introduced dual-attention blocks (DA-Blocks), as depicted in Figure \ref{Fig_DATrans-Unet}, in each of the three skip connection layers. 
This decision was based on our observation that traditional skip connections often transmit redundant features, which DA-Blocks effectively filter.
Integrating DA-Blocks into the skip connections allows them to refine the sparsely encoded features from both positional and channel perspectives, extracting more valuable information while reducing redundancy. 
By doing so, DA-Blocks assist the decoder in more accurate feature map reconstruction. Moreover, the inclusion of DA-Blocks not only enhances the model's robustness but also effectively mitigates sensitivity to overfitting, contributing to the overall performance and generalization capability of the model.
%与其他 U 结构模型类似，我们还在编码器和解码器之间引入了跳跃连接，以弥合它们之间存在的语义差距。 为了进一步缩小这种语义差距，我们在三个跳跃连接层中的每一个层中引入了双重注意块（Da-Blocks），如图 \ref{Fig_DATrans-Unet} 所示。 将 Da-Blocks 集成到跳跃连接中，使他们能够从位置和通道的角度细化稀疏编码的特征，提取更有价值的信息，同时减少冗余。 这反过来又极大地方便了解码器重建原始特征图。 此外，Da-Blocks的加入不仅增强了模型的鲁棒性，还有效降低了过拟合的敏感性，有助于提高模型的整体性能和泛化能力。
\subsection{Decoder}
% HDmm DSC是百分号

%decoder中的两个段落的逻辑需要更加紧密一些, 等等解释一下为什么这样安排, 还需要说一下为什么decoder要这样, 这样的好处也是decoder的重点. 
%TODO: 需要补充原理!  先说decode的作用是和sip配合进行还原, 再说为什么还原.再详细说结构. over
As depicted in Figure \ref{Fig_DATrans-Unet}, the right half of the diagram corresponds to the decoder. The primary role of the decoder is to reconstruct the original feature map by utilizing features acquired from the encoder and those received through skip connections, employing operations like upsampling.

The decoder's components include feature fusion, a segmentation head, and three upsampling convolution blocks. The first component: feature fusion entails the integration of feature maps transmitted through skip connections with the existing feature maps, thereby assisting the decoder in faithfully reconstructing the original feature map. The second component: the segmentation head is responsible for restoring the final output feature map to its original dimensions. The third component: the three upsampling convolution blocks incrementally double the size of the input feature map in each step, effectively restoring the image's resolution.

Putting the above parts together, the workflow begins by passing the input image through convolution blocks and subsequently performing upsampling to augment the size of the feature maps. These feature maps undergo a twofold size increase while their dimensions are reduced by half. The features received through the skip connections are then fused, followed by continued upsampling and convolution. After three iterations of this process, the generated feature map undergoes one final round of upsampling and is accurately restored to its original size by the segmentation head.

Thanks to this architecture, the decoder demonstrates robust decoding capabilities, effectively revitalizing the original feature map using features from both the encoder and skip connections.

\section{Experiments}
\label{experiments}
To evaluate the proposed method, we performed experiments on Synapse\cite{Synapse}, CVC-ClinicDB dataset\cite{cvc-clinicDB}, Chest X-ray mask and label dataset\cite{jaeger2013automatic,candemir2013lung} Analysis, Kvasir SEG dataset\cite{kvasir-Seg}, Kvasir-Instrument dataset\cite{Kvasir-Instrument}, 2018ISIC-Task\cite{ISIC01,ISIC02}. The experimental results demonstrate that DA-TransUNet outperforms existing methods across all six datasets. 
In the following subsections, we first introduce the dataset and implementation details. Then show the results on each of the six datasets.
% Finally, we report experimental results on Synapse, CVC-ClinicDB, Chest X-ray Mask and Tag, Kvasir SEG, Kvasir-Instrument and 2018ISIC-Task.
%为了评估所提出的方法，我们对 Synapse\cite{Synapse}、CVC-ClinicDB 数据集 \cite{cvc-clinicDB}、胸部 X 射线掩模和标签数据集 \cite{jaeger2013automatic} \cite{candemir2013lung} 进行了分析 、Kvasir SEG 数据集\cite{kvasir-Seg}、Kvasir-Instrument 数据集\cite{Kvasir-Instrument}、2018ISIC-Task\cite{ISIC01}\cite{ISIC02}。 实验结果表明 DA-TransUNet 在所有六个数据集上都有所改进。 在接下来的小节中，我们将首先介绍数据集和实现细节。 最后，我们报告了 Synapse、CVC-ClinicDB、胸部 X 射线面罩和标签、Kvasir SEG、Kvasir-Instrument 和 2018ISIC-Task的实验结果。

\subsection{Datasets}

\subsubsection{Synapse}
The Synapse dataset consists of 30 scans of eight abdominal organs. These eight organs include the left kidney, right kidney, aorta, spleen, gallbladder, liver, stomach and pancreas. There are a total of 3779 axially enhanced abdominal clinical CT images. 

\subsubsection{CVC—ClinicDB}
CVC-ClinicDB is a database of frames extracted from colonoscopy videos, which is part of the Endoscopic Vision Challenge. This is a dataset of endoscopic colonoscopy frames for the detection of polyps. CVC-ClinicDB contains 612 still images from 29 different sequences. Each image has its associated manually annotated ground truth covering the polyp.

\subsubsection{Chest Xray}
Chest Xray Masks and Labels X-ray images and corresponding masks are provided. The X-rays were obtained from the Montgomery County Department of Health and Human Services Tuberculosis Control Program, Montgomery County, Maryland, USA. The set of images contains 80 anterior and posterior X-rays, of which 58 X-rays are normal and 1702 X-rays are abnormal with evidence of tuberculosis. All images have been de-identified and presented in DICOM format. The set contains a variety of abnormalities, including exudates and corneal morphology. It contains 138 posterior anterior radiographs, of which 80 radiographs were normal and 58 radiographs showed abnormal manifestations of tuberculosis. 

\subsubsection{Kvasir SEG}
Kvasir SEG is an open-access dataset of gastrointestinal polyp images and corresponding segmentation masks, manually annotated and verified by an experienced gastroenterologist. It contains 1000 polyp images and their corresponding groudtruth, the resolution of the images contained in Kvasir-SEG varies form 332x487 to 1920x1072 pixels, the file format is jpg. 

\subsubsection{Kvasir-Instrument}
Kvasir-Instrument a gastrointestinal instrument Dataset. It contains 590 endoscopic tool images and their groud truth mask, the resolution of the image in the dataset varies from 720x576 to 1280x1024, which consists of 590 annotated frames comprising of GI procedure tools such as snares, balloons, biopsy forceps, etc. the file format is jpg.

\subsubsection{2018ISIC-Task}
The dataset used in the 2018 ISIC Challenge addresses the challenges of skin diseases. It comprises a total of 2512 images, with a file format of JPG. The images of lesions were obtained using various dermatoscopic techniques from different anatomical sites (excluding mucous membranes and nails). These images are sourced from historical samples of patients undergoing skin cancer screening at multiple institutions. Each lesion image contains only a primary lesion.

\subsection{Implementation Settings}

\subsubsection{Baselines}
In our endeavor to innovate in the field of medical image segmentation, we benchmark our proposed model against an array of highly-regarded baselines, including the U-net, UNet++, DA-Unet, Attention U-net, and TransUNet. The U-net has been a foundational model in biomedical image segmentation\cite{U-Net}. Unet++ brings added sophistication with its implementation of intermediate layers\cite{unet++}. The DA-Unet goes a step further by integrating dual attention blocks, amplifying the richness of features extracted\cite{mlda-unet}. The Attention U-net employs an attention mechanism for improved feature map weighting\cite{attention-unet}, and finally, the TransUNet deploys a transformer architecture, setting a new bar in segmentation precision\cite{transunet}. Through this comprehensive comparison with these eminent baselines, we aim to highlight the unique strengths and expansive potential applications of our proposed model. Additionally, we benchmarked our model against advanced state-of-the-art algorithms. UCTansNet allocates skip connections through the attention module in the traditional U-net model\cite{uctransnet}. TransNorm integrates the Transformer module into the encoder and skip connections of standard U-Net \cite{transnorm}. A novel Transformer module was designed and a model named MIM was built with it \cite{MIM}. By extensively comparing our model with current state-of-the-art solutions, we intend to showcase its superior segmentation performance.

\begin{table*}[!htbp]
\caption{\textbf{Experimental results on the Synapse dataset}}
\label{table:Synapse}
\centering
\resizebox{\textwidth}{!}{%
\begin{tabular}{lccccccccccc}
\toprule
Model & Year & DSC$\uparrow$ & HD$\downarrow$ & Aorta & Gallbladder & Kidney(L) & Kidney(R) & Liver & Pancreas & Spleen & Stomach \\
\midrule
U-net\cite{U-Net} & 2015 & 76.85\% & 39.70 & 89.07 & \textbf{69.72} & 77.77 & 68.6 & 93.43 & 53.98 & 86.67 & 75.58 \\
U-Net++\cite{unet++} & 2018 & 76.91\% & 36.93 & 88.19 & 68.89 & 81.76 & 75.27 & 93.01 & 58.20 & 83.44 & 70.52 \\
Residual U-Net\cite{resunet} & 2018 & 76.95\% & 38.44 & 87.06 & 66.05 & \textbf{83.43} & 76.83 & 93.99 & 51.86 & 85.25 & 70.13 \\
Att-Unet\cite{attention-unet} & 2018 & 77.77\% & 36.02 & \textbf{89.55} & 68.88 & 77.98 & 71.11 & 93.57 & 58.04 & 87.30 & 75.75 \\
MultiResUNet\cite{multiresunet} & 2020 & 77.42\% & 36.84 & 87.73 & 65.67 & 82.08 & 70.43 & 93.49 & 60.09 & 85.23 & 75.66 \\
TransUNet\cite{transunet} & 2021 & 77.48\% & 31.69 & 87.23 & 63.13 & 81.87 & 77.02 & 94.08 & 55.86 & 85.08 & 75.62 \\
UCTransNet\cite{uctransnet} & 2022 & 78.23\% & 26.75 & 84.25 & 64.65 & 82.35 & 77.65 & 94.36 & 58.18 & 84.74 & 79.66 \\
TransNorm\cite{transnorm} & 2022 & 78.40\% & 30.25 & 86.23 & 65.1 & 82.18 & 78.63 & 94.22 & 55.34 & 89.50 & 76.01 \\
MIM\cite{MIM} & 2022 & 78.59\% & 26.59 & 87.92 & 64.99 & 81.47 & 77.29 & 93.06 & 59.46 & 87.75 & 76.81 \\
swin-unet\cite{swin-unet} & 2022 & 79.13\% & \textbf{21.55} & 85.47 & 66.53 & 83.28 & 79.61 & 94.29 & 56.58 & \textbf{90.66} & 76.60 \\
\textbf{DA-TransUNet(Ours)} & 2023 & \textbf{79.80\%} & 23.48 & 86.54 & 65.27 & 81.70 & \textbf{80.45} & \textbf{94.57} & \textbf{61.62} & 88.53 & \textbf{79.73} \\
\bottomrule
\end{tabular}%
}
\end{table*}

\begin{figure*}[t!]
\centering
\includegraphics[width=0.7\textwidth]{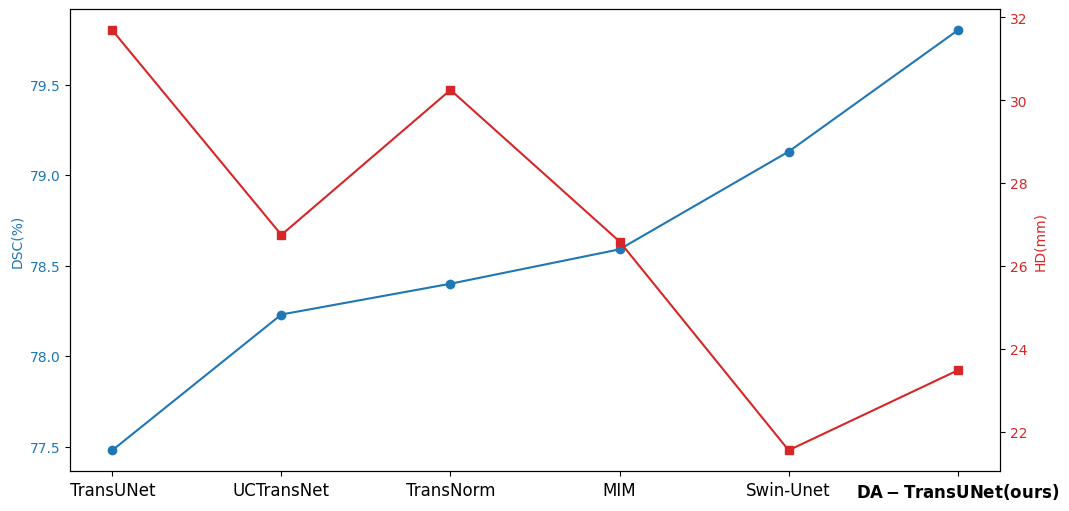}
\caption{\textbf{Line chart of DSC and HD values of several advanced models in the Synapse dataset}}
\label{Fig_line_chart}
\end{figure*}

\begin{figure*}[t!]
\centering
\includegraphics[width=0.6\linewidth]{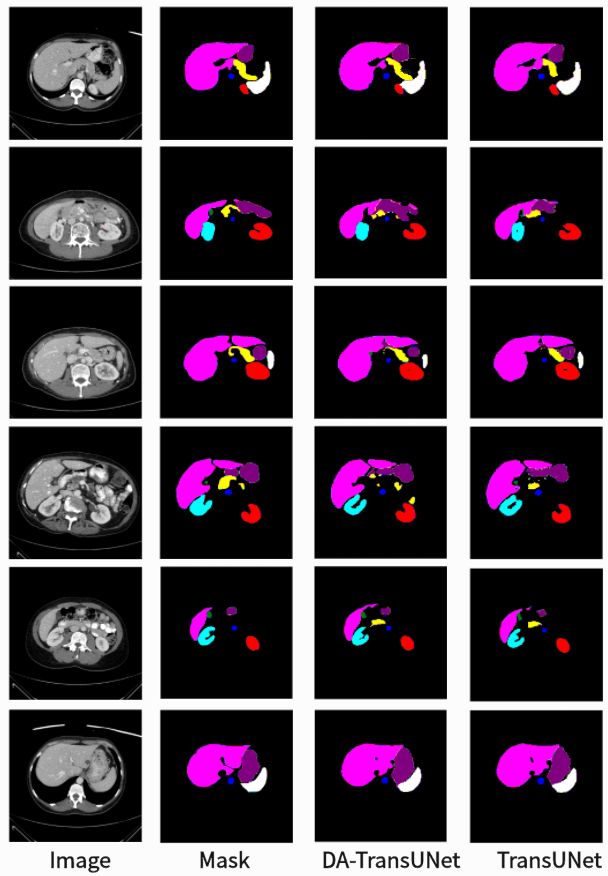}
\caption{\textbf{Segmentation results of TransUNet and DA-TransUNet on the Synapse dataset.}}
\label{Segmentation of Synapse}
\end{figure*}

% \begin{table*}[!ht]
% \caption{Experimental results of datasets (CVC-ClinicDB, Chest Xray Masks and Labels, ISIC2018-Task, kvasir-instrument, kvasir-seg)}
% \label{table:Other}
%     \centering
%     \begin{tabular}{|l|l|l|l|l|l|l|l|l|l|l|}
%     \hline
%         ~ & \multicolumn{2}{|c|}{CVC-ClinicDB}& \multicolumn{2}{|c|}{Chest Xray Masks and Labels} & \multicolumn{2}{|c|}{ISIC2018-Task} & \multicolumn{2}{|c|}{kvasir-instrument}& \multicolumn{2}{|c|}{kvasir-seg}  \\ \hline
%         ~ & Iou↑ & Dice↑ & Iou↑ & Dice↑ & Iou↑ & Dice↑ & Iou↑ & Dice↑ & Iou↑ & Dice↑  \\ \hline
%         Unet & 0.7821 & 0.8693 & 0.9303 & 0.9511 & 0.8114 & 0.8722 & 0.8957 & 0.9358 & 0.8012 & 0.8822  \\ \hline
%         Attn-Unet & 0.7935 & 0.8741 & 0.9274 & 0.9503 & 0.8151 & 0.876 & 0.8949 & 0.9359 & 0.7801 & 0.8661  \\ \hline
%         Unet++ & 0.7847 & 0.8714 & 0.9289 & 0.9505 & 0.8133 & 0.873 & \textbf{0.8995} & \textbf{0.9389} & 0.7767 & 0.8657  \\ \hline
%         ResUNet & 0.5902 & 0.7422 & 0.9262 & 0.9505 & 0.7651 & 0.8332 & 0.8572 & 0.9141 & 0.6604 & 0.7785  \\ \hline
%         TransUNet & 0.8163 & 0.8901 & 0.9301 & 0.9535 & 0.8263 & 0.8878 & 0.8926 & 0.9363 & 0.8003 & 0.8791  \\ \hline
%         DA-TransUNet(Ours) & \textbf{0.8251} & \textbf{0.8947} & \textbf{0.9317} & \textbf{0.9538} & \textbf{0.8278} & \textbf{0.8888} & 0.8973 & 0.9381 & \textbf{0.8102} & \textbf{0.8847}  \\ \hline
%     \end{tabular}
% \end{table*}
\begin{table*}[!htbp]
\caption{\textbf{Experimental results of datasets (CVC-ClinicDB, Chest Xray Masks and Labels, ISIC2018-Task, kvasir-instrument, kvasir-seg)}}
\label{table:Other}
\centering
\resizebox{\textwidth}{!}{%
\begin{tabular}{lccccccccccc}
\toprule
~ & \multicolumn{2}{c}{CVC-ClinicDB} & \multicolumn{2}{c}{Chest Xray Masks and Labels} & \multicolumn{2}{c}{ISIC2018-Task} & \multicolumn{2}{c}{kvasir-instrument} & \multicolumn{2}{c}{kvasir-seg} \\
\cmidrule(lr){2-3} \cmidrule(lr){4-5} \cmidrule(lr){6-7} \cmidrule(lr){8-9} \cmidrule(lr){10-11}
~ & Iou$\uparrow$ & Dice$\uparrow$ & Iou$\uparrow$ & Dice$\uparrow$ & Iou$\uparrow$ & Dice$\uparrow$ & Iou$\uparrow$ & Dice$\uparrow$ & Iou$\uparrow$ & Dice$\uparrow$ \\
\midrule
U-net\cite{U-Net} & 0.7821 & 0.8693 & 0.9303 & 0.9511 & 0.8114 & 0.8722 & 0.8957 & 0.9358 & 0.8012 & 0.8822 \\
Attn-Unet\cite{attention-unet} & 0.7935 & 0.8741 & 0.9274 & 0.9503 & 0.8151 & 0.876 & 0.8949 & 0.9359 & 0.7801 & 0.8661 \\
Unet++\cite{unet++} & 0.7847 & 0.8714 & 0.9289 & 0.9505 & 0.8133 & 0.873 & \textbf{0.8995} & \textbf{0.9389} & 0.7767 & 0.8657 \\
ResUNet\cite{resunet} & 0.5902 & 0.7422 & 0.9262 & 0.9505 & 0.7651 & 0.8332 & 0.8572 & 0.9141 & 0.6604 & 0.7785 \\
TransUNet\cite{transunet} & 0.8163 & 0.8901 & 0.9301 & 0.9535 & 0.8263 & 0.8878 & 0.8926 & 0.9363 & 0.8003 & 0.8791 \\
\textbf{DA-TransUNet(Ours)} & \textbf{0.8251} & \textbf{0.8947} & \textbf{0.9317} & \textbf{0.9538} & \textbf{0.8278} & \textbf{0.8888} & 0.8973 & 0.9381 & \textbf{0.8102} & \textbf{0.8847} \\
\bottomrule
\end{tabular}%
}
\end{table*}

\subsubsection{Implementation Details}
We implemented DA-TransUNet using the PyTorch framework and trained it on a single NVIDIA RTX 3090 GPU \cite{pytorch}.
The model was trained with an image resolution of 256x256 and a patch size of 16. We employed the Adam optimizer, configured with a learning rate of  1e-3, momentum of 0.9, and weight decay of 1e-4. 
All models were trained for 500 epochs unless stated otherwise. In order to ensure the convergence of the indicators, but due to different data set sizes, we used 50 epochs for training on the two data sets, Chest Xray Masks and Labels and ISIC 2018-Task.

During the training phase on five datasets, including CVC-ClinicDB, the proposed DA-TransUNet model is trained in an end-to-end manner. Its objective function consists of a weighted binary cross-entropy loss function (BCE) and a Dice coefficient loss function. To facilitate training, the final loss function, termed "Loss," is formulated as follows:
%在包括 CVC-ClinicDB 在内的五个数据集的训练阶段，所提出的 DA-TransUNet 模型以端到端的方式进行训练。 其目标函数由加权二元交叉熵损失函数（BCE）和 Dice 系数损失函数组成。 为了帮助模型训练，最终的损失函数（表示为“Loss”）可以表示如下：
\begin{equation}
    \text{Loss} = \frac{1}{2} \times \text{BCE} + \frac{1}{2} 
    \times \text{DiceLoss}
\end{equation}

To ensure a fair evaluation of the Synapse dataset, we utilized the pre-trained model "R50-ViT" with input resolution and patch size set to 224x224 and 16, respectively. We trained the model using the SGD optimizer, setting the learning rate to 0.01, momentum of 0.9, and weight decay of 1e-4. The default batch size was set to 24. The loss function employed for the Synapse dataset is defined as follows:
\begin{equation}
    \text{Loss} = \frac{1}{2} \times \text{Cross-Entropy Loss} + \frac{1}{2} 
    \times \text{DiceLoss}
\end{equation}

This loss function balances the contributions of cross-entropy and Dice losses, ensuring impartial evaluation during testing on the Synapse dataset.
%TODO: 详细介绍损失函数, 并给出公式 over

When using the datasets, we use a 3 to 1 ratio, where 75\% is the training set and 25\% is the test set, to ensure adequacy of training.

\subsubsection{Model Evaluation}
% Let us first introduce the three evaluation criteria used.
In evaluating the performance of DA-TransUNet, we utilize a comprehensive set of metrics including Intersection over Union (IoU), Dice Coefficient(DSC), and Hausdorff Distance (HD). These metrics are industry standards in computer vision and medical image segmentation, providing a multifaceted assessment of the model's accuracy, precision, and robustness.

IOU (Intersection over Union) is one of the commonly used metrics to evaluate the performance of computer vision tasks such as object detection, image segmentation and instance segmentation. It measures the degree of overlap between the predicted region of the model and the actual target region, which helps us to understand the accuracy and precision of the model. In target detection tasks, IOU is usually used to determine the degree of overlap between the predicted bounding box (Bounding Box) and the real bounding box. In image segmentation and instance segmentation tasks, IOU is used to evaluate the degree of overlap between the predicted region and the ground truth segmentation region.
%IOU（Intersection over Union）是评估物体检测、图像分割和实例分割等计算机视觉任务性能的常用指标之一。它衡量的是模型预测区域与实际目标区域的重叠程度，有助于我们了解模型的准确性和精确度。在目标检测任务中，IOU 通常用于判断预测边界框（Bounding Box）与真实边界框之间的重叠程度。在图像分割和实例分割任务中，IOU 用于评估预测区域和地面实况分割区域之间的重叠程度。

\begin{equation}I O U=\frac{T P}{F P+T P+F N}\label{iou}\end{equation}

The Dice coefficient (also known as the Sørensen-Dice coefficient, F1-score, DSC) is a measure of model performance in image segmentation tasks, and is particularly useful for dealing with class imbalance problems. It measures the degree of overlap between the predicted results and the ground truth segmentation results, and is particularly effective when dealing with segmentation of objects with unclear boundaries. The Dice coefficient is commonly used as a measure of the model's accuracy on the target region in image segmentation tasks, and is particularly suitable for dealing with relatively small or uneven target regions.
%Dice 系数（又称 Sørensen-Dice 系数、F1-分数）是用于衡量图像分割任务中模型性能的指标，尤其适用于处理类不平衡问题。它衡量的是预测结果与地面实况分割结果之间的重叠程度，在处理边界不清晰的物体分割时尤其有效。在图像分割任务中，Dice 系数通常用来衡量模型对目标区域的准确性，尤其适用于处理相对较小或不平整的目标区域。

\begin{equation}
        \operatorname{Dice}(P, T) = \frac{| P_{1} \cap T_{1} |}{| P_{1} | + | T_{1} |} \Leftrightarrow \text{Dice} = \frac{2 | T \cap P |}{| F | + | P |}
\label{ppv}\end{equation}

Hausdorff Distance (HD) is a distance measure for measuring the similarity between two sets and is commonly used to evaluate the performance of models in image segmentation tasks. It is particularly useful in the field of medical image segmentation to quantify the difference between predicted and true segmentations. The computation of Hausdorff distance captures the maximum difference between the true segmentation result and the predicted segmentation result, and is particularly suitable for evaluating the performance of segmentation models in boundary regions.
%豪斯多夫距离（Hausdorff Distance，简称 HD）是一种用于测量两个集合之间相似性的距离测量方法，通常用于评估图像分割任务中模型的性能。它在医学影像分割领域尤其有用，可用于量化预测分割与真实分割之间的差异。豪斯多夫距离的计算方法可以捕捉真实分割结果与预测分割结果之间的最大差异，尤其适用于评估分割模型在边界区域的性能。
\begin{equation}
H(A, B)=\max \left\{\max _{a \in A} \min _{b \in B}\|a-b\|, \max _{b \in B} \min _{a \in A}\|b-a\|\right\}
\label{dice}\end{equation}

We evaluate using both Dice and HD in the Synapse dataset and both Dice and IOU in other datasets.
%我们在Synapse数据集中使用看Dice和HD两者，在其他数据集中使用Dice和IOU两者进行评估。

\begin{figure*}[t!]
\centering
\includegraphics[width=\textwidth]{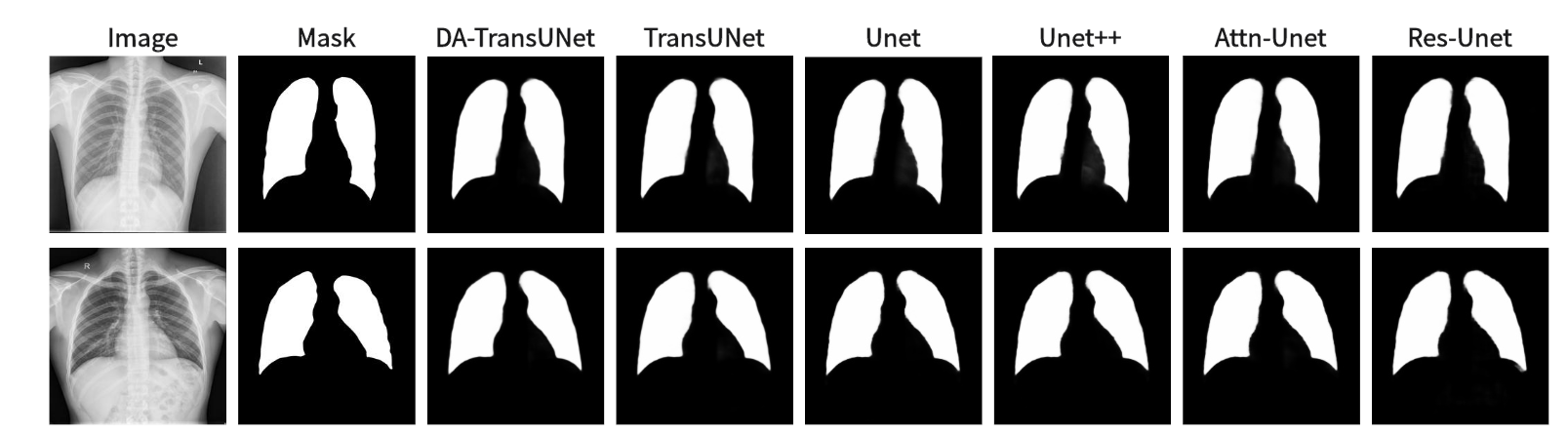}
\caption{\textbf{Comparison of qualitative results between DA-TransUNet and existing models on the task of segmenting Chest X-ray Masks and Labels X-ray datasets.}}
\label{Segmentation of Chest}
\end{figure*}

\begin{figure*}[t!]
\centering
\includegraphics[width=\textwidth]{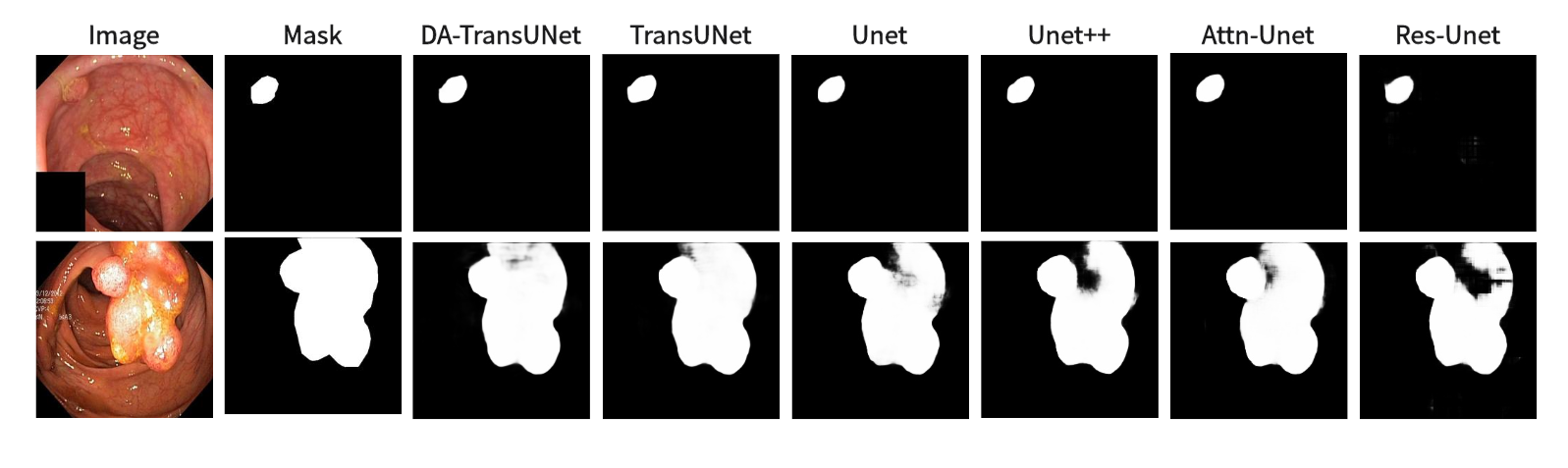}
\caption{\textbf{Comparison of qualitative results between DA-TransUNet and existing models on the task of segmenting Kvasir-Seg datasets.}}
\label{Segmentation of seg}
\end{figure*}

\begin{figure*}[t!]
\centering
\includegraphics[width=\textwidth]{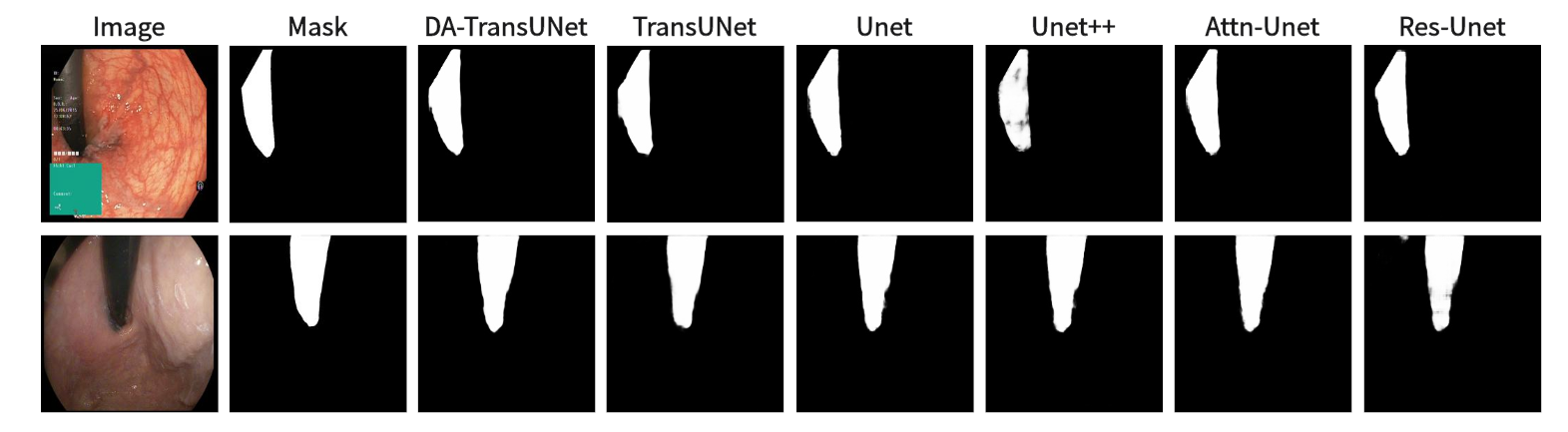}
\caption{\textbf{Comparison of qualitative results between DA-TransUNet and existing models on the task of segmenting Kavsir-Instrument datasets.}}
\label{Segmentation of instrument}
\end{figure*}

\begin{figure*}[t!]
\centering
\includegraphics[width=\textwidth]{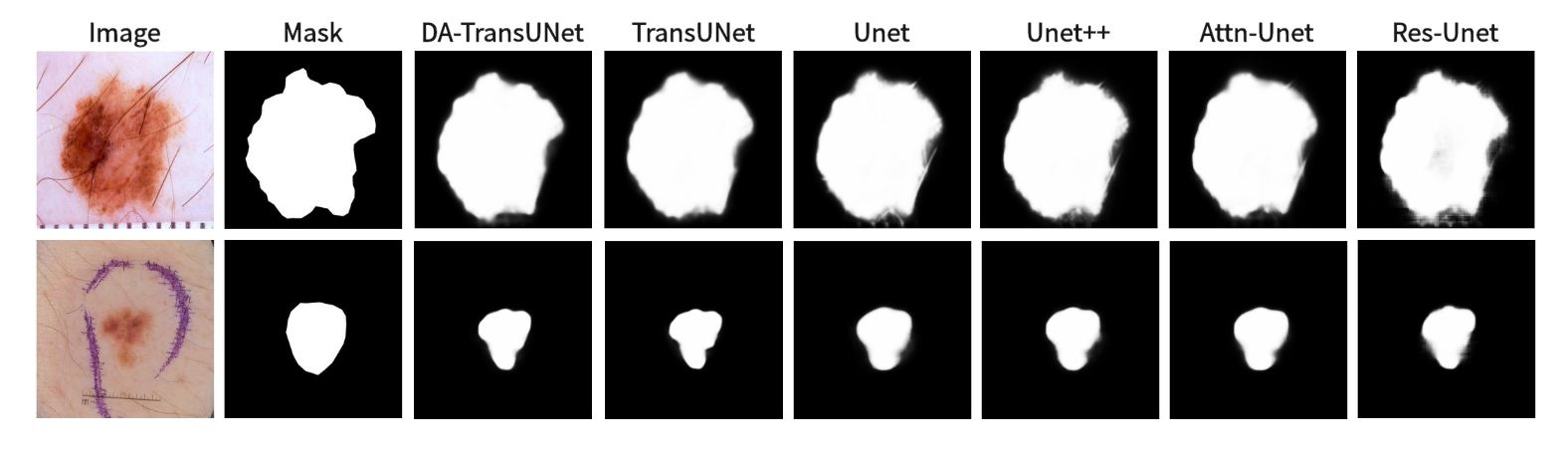}
\caption{\textbf{Comparison of qualitative results between DA-TransUNet and existing models on the task of segmenting 2018ISIC-Task datasets.}}
\label{Segmentation of ISIC}
\end{figure*}

\begin{figure*}[t!]
\centering
\includegraphics[width=\textwidth]{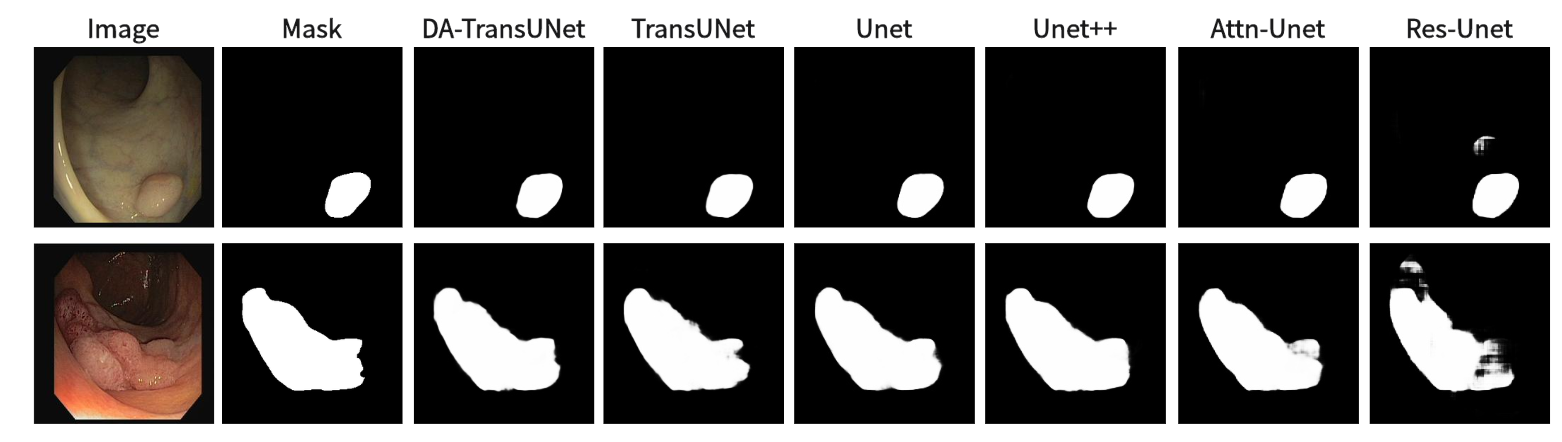}
\caption{\textbf{Comparison of qualitative results between DA-TransUNet and existing models on the task of segmenting CVC-ClinicDB datasets.}}
\label{Segmentation of CVC}
\end{figure*}

\subsection{Comparison to the State-of-the-Art Methods}

We have chosen U-net\cite{U-Net}, Res-Unet\cite{resunet}, TransUNet\cite{transunet}, U-Net++\cite{unet++}, Att-Unet\cite{attention-unet}, TransNorm\cite{transnorm},  UCTransNet\cite{uctransnet}, MultiResUNet\cite{multiresunet}, swin-unet\cite{swin-unet} and MIM\cite{MIM} to compare with our DA-TransUNet, and the experimental data are tabulated below.

In order to demonstrate the superiority of the DA-TransUNet model proposed in this paper, we conducted the main experiments using the Synapse dataset and compared it with its 11 state-of-the-art models (SOTA) (see Table\ref{table:Synapse}). 

As shown in the Figure\ref{Segmentation of Synapse}, we can see that the average DSC and average HD evaluation criteria are 79.80\% and 23.48 mm, respectively, which are improved by 2.32\% and 8.21 mm, respectively, compared with TransUNet, which indicates that our DA-TransUNet has better segmentation ability than TransUNer in terms of overall segmentation results and organ edge prediction. As shown in the Figure\ref{Fig_line_chart}, on the other hand, we can see that DSC has the highest value of our model. Although HD is higher than Swin-Unet, it is still an improvement compared to several newer models and TransUNet. The segmentation time for an image is 35.98 ms for our DA-TransUNet and 33.58 ms for TransUNet, which indicates that there is not much difference in the segmentation speed between the two models, but our DA-TransUNet has better segmentation results. In the segmentation results of 8 organs, DA-TransUNet outperforms TransUNet by 2.14\%, 3.43\%, 0.48\%, 3.45\%, and 4.11\% for the five datasets of Gallbladder, right kidney, liver, spleen, and stomach, respectively. The segmentation rate for the pancreas is notably higher at 5.73\%.
In a comparative evaluation across six distinct organs, DA-TransUNet demonstrates superior segmentation capabilities relative to TransUNet. Nevertheless, it exhibits a marginal decrement in the segmentation accuracy for the aorta and left kidney by 0.69\% and 0.17\%, respectively. The model achieves the best segmentation rates for the right kidney, liver, pancreas, and stomach, indicating superior feature learning capabilities on these organs.

%为了证明本文提出的 DA-TransUNet 模型的优越性，我们利用 Synapse 数据集进行了主要实验，并与其 10 个最先进的模型（SOTA）进行了比较（见表 1）。另一方面，如图\ref{Fig_line_chart}所示，我们可以看到 DSC 具有我们模型的最高值。 虽然HD比Swin-Unet要高，但与几个较新的模型和TransUNet相比仍然是一个进步。从表可见，平均DSC和平均HD评价标准分别为80.11%和26.38 mm，与TransUNet相比分别提高了2.32%和8.21 mm，这表明我们的DA-TransUNet在整体分割结果和器官边缘预测方面比TransUNer具有更好的分割能力。我们的 DA-TransUNet 对一幅图像的分割时间为 35.98 毫秒，TransUNet 为 33.58 毫秒，这表明两种模型的分割速度相差不大，但我们的 DA-TransUNet 的分割结果更为出色。在 8 个器官的分割结果中，DA-TransUNet 在胆囊，右肾，肝脏，脾，胃，这五个数据集分别比TransUnet高出了2.14\%，3.43\%，0.48\%，3.45\%，4.11\%。其中胰腺的分割率更是达到了惊人的5.73\%。可见DA-TransUNet在这六个器官中的分割能力均强于TransUNet。可惜在主动脉和胆囊的分割中比TransUNet低了0.69\%，0.17\%。同时，在所有模型中右肾、肝脏、胰腺、胃这四个器官的分割成功率达到了最优，可以得出我们的模型在这四个器官上的特征学习能力最优秀。

To further confirm the better segmentation of our model compared to TransUNet, we visualized the segmentation plots of TransUNet and DA-TransUNet (see Figure\ref{Segmentation of Synapse}). From the yellow and purple parts in the first column, we can see that our segmentation effect is obviously better than that of TransUNet; from the second column, the extension of purple is better than that of TransUNet, and there is no vacancy in the blue part; from the third column, there is a semicircle in the yellow part, and the vacancy in red is smaller than that of TransUNet, etc. It is evident that DA-TransUNet outperforms TransUNet in segmentation quality. In summary, DA-TransUNet significantly surpasses TransUNet in segmenting the left kidney, right kidney, spleen, stomach, and pancreas. It also offers superior visualization performance in image segmentation.
%为了进一步证实我们模型较于TransUNet的分割效果更好，我们可视化了TransUNet和DA-TransUNet的分割图（see Figure）。从第一列图中黄色和紫色部分可见我们分割效果明显优于TransUNet；从第二列中紫色延展性比TransUNet好，蓝色部分没有空缺；第三列的黄色部分的有半圆，红色空缺比TransUNet小等等。综合来看可以得出，我们模型对于左肾，右肾，脾，胃和胰腺的分割结果明显优于TransUNet。可以得出，DA-TransUNet在图像分割可视化中的表现也是优于TransUNet。

We simultaneously took DA-TransUNet in five datasets, CVC-ClinicDB, Chest Xray Masks and Labels, ISIC2018-Task, kvasir-instrument, and kvasir-seg, and compared it with some classical models (see Table \ref{table:Other}). In the table, the values of IOU and Dice of DA-TransUNet are higher than TransUNet in all five datasets, CVC-ClinicDB, Chest Xray Masks and Labels, ISIC2018-Task, kvasir-instrument, and kvasir-seg. Also DA-TransUNet has the best dataset segmentation in four of the five datasets. As seen in the table, our DA-TransUNet has more excellent feature learning and image segmentation capabilities.
%我们同时在CVC-ClinicDB，Chest Xray Masks and Labels，ISIC2018-Task，kvasir-instrument，kvasir-seg这五个数据集中拿DA-TransUNet与一些经典模型进行了对比（see table 2）。在表中，DA-TransUNet的IOU和Dice数值在CVC-ClinicDB，Chest Xray Masks and Labels，ISIC2018-Task，kvasir-instrument，kvasir-seg这五个数据集中都比TransUNet高。同时DA-TransUNet在五个数据集中四个数据集分割效果最好。从表中可见，我们的DA-TransUNet有着更加优秀的特征学习和图像分割能力。

We also show the results of image segmentation visualization of DA-TransUNet in these five datasets, and we also show the results of the comparison models for the comparison. 
The visualization results for Chest X-ray Masks and Labels, Kvasir-Seg, Kvasir-Instrument, ISIC2018-Task, and CVC-ClinicDB datasets are presented in Figure\ref{Segmentation of Chest}, Figure\ref{Segmentation of seg}, Figure\ref{Segmentation of instrument}, Figure\ref{Segmentation of ISIC}, and Figure\ref{Segmentation of CVC}, respectively.
In the Figure, it can be seen that the segmentation effect of DA-TransUNet has a good performance.
Firstly, DA-TransUNet has better segmentation results than TransUNet. In addition, compared with the four classical models of U-net, Unet++, Attn-Unet, and Res-Unet, DA-TransUNet has a certain improvement. It can be seen that the effectiveness of DA-TransUNet for model segmentation is not only confirmed in the Synapse dataset, but also in the five datasets (CVC-ClinicDB, Chest Xray Masks and Labels, ISIC2018-Task, kvasir-instrument, kvasir-seg). We further establish that DA-TransUNet excels in both 3D and 2D medical image segmentation.
%我们也将DA-TransUNet在这五个数据集中图像分割可视化的结果展现了出来，同时对了对比我们也将对比模型的结果进行了展现。首先DA-TransUNet较于TransUNet有着更加出色得分割结果。再者与Unet，Unet++，Attn-Unet，Res-Unet这四个经典模型相比较又有一定的提升。可见，DA-TransUNet对于模型分割的有效性不仅在Synapse数据集中得到了证实，同时也在这五个数据集（CVC-ClinicDB, Chest Xray Masks and Labels, ISIC2018-Task, kvasir-instrument, kvasir-seg）中得到了证明。同时我们也证明了DA-TransUNet不仅只是在3D医学图像分割中有着提升，在2D医学图像分割中也有提升。

% \begin{figure*}[t!]
% \centering
% $\in$cludegraphics[width=0.7\linewidth]{images/Synapse.png}
% \caption{\textbf{Segmentation results of TransUNet and DA-TransUNet on the Synapse dataset.}}
% \label{Segmentation of Synapse}
% \end{figure*}

% \begin{figure*}[t!]
% \centering
% $\in$cludegraphics[width=\textwidth]{images/other.png}
% \caption{\textbf{Segmentation results for several classical models in the dataset.}}
% \label{Segmentation of other}
% \end{figure*}

% \begin{table}[!ht]
% \caption{The results of the ablation experiment(Two empty numbers one means yes, zero means no. The first blank represents the skip connection layer, and the latter layer represents the encoding layer.)}
% \label{table:Ablation01}
%     \centering
%     \begin{tabular}{|l|l|l|}
%     \hline
%         ~ & DSC↑  & HD↓  \\ \hline
%         DA-TransUNet(1$\backslash$0) & 77.9 & 26.47  \\ \hline
%         DA-TransUNet(0$\backslash$1) & 79.13 & 34.32  \\ \hline
%         DA-TransUNet(0$\backslash$0) & 77.48 & 31.69  \\ \hline
%         DA-TransUNet(1$\backslash$1) & \textbf{80.11} & \textbf{26.38}  \\ \hline
%     \end{tabular}
% \end{table}
\begin{table}[!htbp]
\caption{\textbf{Effects of Combinatorial Placement of DA-Blocks in the Encoder and Through Skip Connections on Performance Metrics}}
\label{table:Ablation01}
\centering
\begin{tabular}{ccccc}
\hline
~ & Encoder with DA & Skip with DA & DSC$\uparrow$ & HD$\downarrow$ \\
\hline
DA-TransUNet & & & 77.48 & 31.69 \\
DA-TransUNet & & $\surd $ & 78.28 & 29.09 \\
DA-TransUNet & $\surd $ & & 78.87 & 27.71 \\
DA-TransUNet & $\surd $ & $\surd $ & \textbf{79.80} & \textbf{23.48} \\
\hline
\end{tabular}
\end{table}

\begin{table}[!ht]
\caption{\textbf{Effects of Incorporating DA-Block in the Encoder and Skip Connections at Different Layers on Performance Metrics}}
\label{table:Ablation02}
    \centering
    \begin{tabular}{cccccc}
    \hline
        ~ & 1st layer & 2nd layer & 3rd layer & DSC↑  & HD↓  \\ 
        \hline
        DA-TransUNet& & & & 78.87 & 27.71  \\ 
        DA-TransUNet&$\surd $ & & & 79.36 & 25.80  \\ 
        DA-TransUNet& & $\surd $& & 78.65 & 23.43  \\ 
        DA-TransUNet& & & $\surd $& 79.49 & 30.71  \\ 
        DA-TransUNet&$\surd $ &$\surd $ & $\surd $& \textbf{79.80} & \textbf{23.48}  \\ \hline
    \end{tabular}
\end{table}

% \begin{table}[!ht]
% \caption{\textbf{Results of the effect of adding feature extraction blocks by skipping connections in different layers}}
% \label{table:Ablation02}
%     \centering
%     \begin{tabular}{ccccccc}
%     \hline
%         ~ & Encoder wieth DA & 1st layer & 2nd layer & 3rd layer & DSC↑  & HD↓  \\ 
%         \hline
%         DA-TransUNet& $\surd $ & & & & 78.87 & 27.71  \\ 
%         DA-TransUNet& $\surd $ & $\surd $ & & & 79.36 & 25.80  \\ 
%         DA-TransUNet& $\surd $ & & $\surd $& & 78.65 & 23.43  \\ 
%         DA-TransUNet& $\surd $ & & & $\surd $& 79.49 & 30.71  \\ 
%         DA-TransUNet& $\surd $ & $\surd $ &$\surd $ & $\surd $& \textbf{79.80} & \textbf{23.48}  \\ \hline
%     \end{tabular}
% \end{table}

\subsection{Ablation Study}
\label{ablation}
We conducted ablation experiments on the DA-TransUNet model using the Synapse dataset to discuss the effects of different factors on model performance. Specifically, it includes: 1) DA-Block in Encoder. 2) DA-Block in Skip Connection.

\subsubsection{Effect of the DA-Block in Encoder And Skip Connection}

In this research (see Table \ref{table:Ablation01}), we conducted experiments to assess the impact of integrating DA-Blocks into the encoder and skip connections on the model's segmentation performance. To be specific, we introduced DA-Blocks into each layer of the skip connections. The results demonstrated a noteworthy improvement: the DSC baseline saw an increase from 77.48\% to 78.28\%, HD index dropped from 31.69mm to 29.09mm. This indicates that the addition of DA-Blocks at each skip connection layer provided the decoder with more refined features, mitigating feature loss during the upsampling process, thereby reducing the risk of overfitting and enhancing model stability. Furthermore, incorporating DA-Blocks into the encoder before the Transformer yielded a significant enhancement, with the DSC baseline increasing from 77.48\% to 78.87\%, even though the HD metric decreased from 31.69mm to 27.71mm. In conclusion, based on the findings presented in Table \ref{table:Ablation01}, we can assert that the inclusion of DA-Blocks both before the Transformer layer and within the skip connections effectively boosts medical image segmentation capabilities.

\subsubsection{Effect of adding DA-Blocks to skip connections in different layers} %SUN TODO: 按照具体实验结果, 规律, 分析原因, 结论推广.重写这个段落-> Finished
Building on the quantitative results from Table \ref{table:Ablation02}, we experimented with various configurations of DA-Block placement across three different layers of skip connections to identify the optimal architectural layout for enhancing the model's performance. Specifically, when DA-Blocks were added to just the first layer, the DSC metric improved to 79.36\% from a baseline of 78.87\%, and the HD metric decreased to 25.80mm from 27.71mm. The addition of DA-Blocks to the second and third layers also showed similar improvements, but the most significant enhancement was observed when DA-Blocks were integrated across all layers, yielding a DSC of 79.80\% and an HD of 23.48mm.
In contrast to traditional architectures where skip connections indiscriminately pass features from the encoder to the decoder, our approach with DA-Blocks selectively improves feature quality at each layer. The results, as corroborated by Table \ref{table:Ablation02}, reveal that introducing DA-Blocks to even a single layer enhances performance, and the greatest gains are observed when applied across all layers. This indicates the effectiveness of integrating DA-Blocks within skip connections for enhancing both feature extraction and medical image segmentation.
Therefore, the table clearly supports the idea that layer-wise inclusion of DA-Blocks in skip connections is an effective strategy for enhancing medical image segmentation.
% As shown in Table \ref{table:Ablation02} DA-TransUNet (the first layer has DA blocks), DA-TransUNet (the second layer has DA blocks), DA-TransUNet (the third layer has DA block), DA-TransUNet(the first, second and third layers all have DA blocks), it can be seen that adding a DA-Block to any of the layers of the skip-connections has a significant improvement in the segmentation of the model, on top of adding a DA-block to the coding layer. 
% As shown in Table \ref{table:Ablation02}, it can be seen that adding a DA-Block to any of the layers of the skip-connections has a significant improvement in the segmentation of the model.
% Based on the table\ref{table:Ablation02}, it can be seen that adding DA-Block in every layer of skip connection is the best effect. 
% The traditional hopping connection only delivers the features from the encoding layer without considering the quality of the delivered features. According to the experiments, it can be concluded that delivering more accurate features to the decoding layer can improve the feature learning ability and image segmentation ability of the model to a greater extent.
%在本研究中，我们测试了在哪几层跳跃连接中加入DA-Block能实现模型分割率的最大提升。如表中DA-TransUNet(1-da),DA-TransUNet(2-da),DA-TransUNet(3-da),DA-TransUNet(1\&0),相比较，可以的见，再编码层加入DA块的基础上，在任意的一层跳跃连接中加入DA-Block对于模型的分割率都有着提升。再根据DA-TransUNet(1\&1)可见在每一层跳跃连接中都加入DA-Block是效果最好的。传统的跳跃连接只是将编码层的特征进行输送，并没有考虑到输送特征的品质，根据实验可以得出输送更准确的的特征给解码层，能够更大程度的提高模型的特征学习能力与图像分割能力。

\section{Discussion}
In this present study, we have discovered promising outcomes from the integration of DA-Blocks with the Transformer and their combination with skip-connections. Encouraging results were consistently achieved across all six experimental datasets.
%TODO: SUN整体改 ->Finished

To start with, drawing from empirical results in Table \ref{table:Ablation01}, it is demonstrated that the integration of DA-Block within the encoder significantly enhances the feature extraction capabilities as well as its segmentation performance. 
In the landscape of computer vision, Vision Transformer (ViT) has been lauded for its robust global feature extraction capabilities \cite{ViT}. However, its falls short in specialized tasks like medical image segmentation, where attention to image-specific features is crucial. To remedy this, in DA-TransUNet we strategically place DA-Blocks ahead of the Transformer module. These DA-Blocks are tailored to first extract and filter image-specific features, such as spatial positioning and channel attributes. Following this initial feature refinement, the processed data is then fed into the Transformer for enhanced global feature extraction. This approach results in significantly improved feature learning and segmentation performance.
In summary, the strategic placement of DA-Blocks prior to the Transformer layer constitutes a pioneering approach that significantly elevates both feature extraction efficacy and medical image segmentation precision.
% In the field of computer vision, ViT has successfully put Transformer to use and achieved impressive results \cite{ViT}. This proves the powerful feature extraction capability of the Transformer. 
% In DA-TransUNet, combining dual-attention blocks with the Transformer demonstrates even better feature learning capabilities. 
% Before the Transformer layer, we introduced DA-blocks to provide more accurate features, which further enhanced the feature extraction capability of Transformer. 
% Our experiments show that good results can be achieved by combining only the dual attention block with the Transformer (see Table\ref{table:Ablation01}), which suggests that the dual attention block can further enhance the global feature extraction capability of the Transformer without creating redundancy.

Morever, building on empirical data in Table \ref{table:Ablation02}, our integration of DA-Blocks with skip connections significantly improves semantic continuity and the decoder's ability to reconstruct accurate feature maps. While traditional U-Net architectures \cite{U-Net} utilize skip connections to bridge the semantic gap between encoder and decoder, our novel incorporation of Dual Attention Blocks within the skip-connection layers yields promising results.
By incorporating DA-Blocks across skip-connection layers, we focus on relevant features and filter out extraneous information, making the image reconstruction process more efficient and accurate.
In summary, the strategic inclusion of DA-Blocks in skip connections represents a groundbreaking approach that not only enhances feature extraction but also improves the model's performance in medical image segmentation.
% Moreover, fusing DA-Block with skip connections can improves semantic bridging and aids feature map recovery by the decoder. In the traditional U-net\cite{U-Net} architecture, skip connections play a role in bridging the semantic gap between encoder and decoder. Initially, researchers successfully combined dual attention blocks and residual blocks into the first skip connection layer with promising results \cite{daresunet}. 
% In our experiments, we introduced dual attention blocks only in the first skip-connection layer with good results (see Table\ref{table:Ablation02} ), thus confirming the effective feature extraction capability of skip-connections. However, we believe that adding dual attention blocks in only one layer is not sufficient to maximize the enhancement of the model. Further integration of dual-attention blocks into the second and third skip-connection layers resulted in notable performance gains at each layer. Refining features before decoding enhances information retention and improves image quality by removing extraneous details.

Despite the advantages, our model also has some limitations.
Firstly, the introduction of the DA-Blocks contributes to an increase in computational complexity. This added cost could potentially be a hindrance in real-time or resource-constrained applications.
Secondly, the decoder part of our model retains the original U-Net architecture. While this design choice preserves some of the advantages of U-Net, it also means that the decoder has not been specifically optimized for our application. This leaves room for further research and improvements, particularly in the decoder section of the architecture.
%在本研究中，我们发现了双注意力（DA）机制块与 Transformer 的集成以及它们与跳跃连接的组合的有希望的结果。 在Synapse数据集上，Dice系数和Hausdorff距离（HD）分别达到80.11%和26.38mm。 此外，在另外五个数据集上也取得了良好的结果。
% 首先，Dual Attention Block (DA-Block)与Transformer的结合可以有效提高Transformer的特征提取能力和全局信息保留能力。 在计算机视觉领域，ViT已经成功使用Transformer并取得了令人瞩目的成果。 这证明了Transformer强大的特征提取能力。 在我们的实验中，将双注意力模块与 Transformer 相结合展示了更好的特征学习能力。 在Transformer层之前，我们引入了DA块来提供更准确的特征，这进一步增强了Transformer的特征提取能力。 我们的实验表明，仅将双重注意块与 Transformer 相结合就可以取得良好的结果（参见 Table\ref{table:Ablation01}），这表明双重注意块可以进一步增强 Transformer 的全局特征提取能力，而无需 造成冗余。
% 此外，将 DA-Block 与跳跃连接融合可以增强跳跃连接的功效，以弥合语义间隙并帮助解码器恢复特征图。 在传统的 Unet\cite{U-Net} 架构中，跳跃连接起到了弥合编码器和解码器之间语义差距的作用。 最初，研究人员成功地将双重注意力块和残差块组合到第一个跳跃连接层中，并取得了有希望的结果/引用{daresunet}。 在我们的实验中，我们仅在第一个跳跃连接层引入双重注意块，并取得了良好的结果（见表），从而证实了跳跃连接的有效特征提取能力。 然而，我们认为仅在一层中添加双重注意块不足以最大化模型的增强。 我们进一步将双注意力块集成到跳跃连接的第二层和第三层中，并观察到集成到每一层中的值得称赞的性能。 显然，在将保留的编码特征输入解码层之前对特征进行细化可以增强信息保留，并通过消除无关细节来改善图像恢复。
% 尽管有这些优点，我们的模型也有一些局限性。
% 首先，双重注意力（DA）块的引入导致计算复杂度的增加。 这种增加的成本可能会成为实时或资源受限应用程序的障碍。
% 其次，我们模型的解码器部分保留了原始的 U-Net 架构。 虽然这种设计选择保留了 U-Net 的一些优点，但这也意味着解码器尚未针对我们的应用进行专门优化。 这为进一步研究和改进留下了空间，特别是在架构的解码器部分。
\section{Conclusion}
% In this paper, we innovatively propose a DA-TransUNet model. To the best of our knowledge, this is the first study that combines DA blocks with Transformer for medical image segmentation tasks. By introducing DA blocks, the model provides more accurate features to the Transformer module in the coding layer, which improves the feature learning ability of the model and increases the segmentation capability of the model. We also incorporate the DA block into the skip connection layer, which reduces the semantic divide that exists in the traditional U-shaped encoder-decoder structure, thus improving the robustness of the model and reducing the risk of overfitting. Experimental results on six public datasets show that our proposed DA-TransUNet has good medical image segmentation capability. In addition, we believe that a large-scale replacement of CNNs in the U-shaped structure with Transformer modules, as in swin-unet, does not necessarily yield good results, and that a proper focus should be placed on the Transformer's own characteristics to improve the feature extraction capability.
In this paper, we innovatively proposed a novel approach to image segmentation by integrating DA-Blocks with the Transformer in the architecture of TransUNet. The DA-Blocks, focusing on image-specific position and channel features, were further integrated into the skip connections to enhance the model's performance. Our experimental results, validated by an extensive ablation study, showed significant improvements in the model's performance across various datasets, particularly the Synapse dataset.

Our research revealed the potential of DA-Block in enhancing the feature extraction capability and global information retention of the Transformer. The integration of DA-Block and Transformer substantially improved the model's performance without creating redundancy. Furthermore, the introduction of DA-Blocks into skip connections not only effectively bridges the semantic gap between the encoder and decoder, but also refines the feature maps, leading to an enhanced image segmentation performance.

% Yet, our model is not without its limitations. The increased computational complexity due to the addition of DA blocks may pose challenges in real-time or resource-constrained applications. The decoder part of our model, following the original U-Net architecture, remains an area for potential optimization and further research.

This study has paved the way for the further use of DA-block in the field of image segmentation. Future work may focus on optimizing the decoder part of our architecture and exploring methods to reduce the computational complexity introduced by DA blocks without compromising the model's performance. We believe our approach can inspire future research in the domain of medical image segmentation and beyond.

\section*{Acknowledgments}
% This was was supported in part by......

%Bibliography
\bibliographystyle{unsrt}  
\bibliography{references}

\end{document}